%% file: ijnme.tex
\pdfoutput=1
\newif\ifnme
\nmefalse

\ifnme

\else
\documentclass{article}

\usepackage{times}

\usepackage{graphicx}
\usepackage{pifont,latexsym,ifthen,rotating,calc,textcase,booktabs,color}
\usepackage{amsfonts,amssymb,amsbsy,amsmath,amsthm}

\renewcommand{\footnotesize}{\fontsize{8.5}{9pt}\selectfont}
\renewcommand{\scriptsize}{\fontsize{8.5}{9.5pt}\selectfont}

\usepackage{fullpage,fancyhdr}
\usepackage[rm,tiny,center,uppercase]{titlesec}
\titlelabel{\thetitle.\quad}
\titleformat{\subsection}[hang]
{\normalfont\itshape}{\thesubsection.}{6pt}{}

\fi

\usepackage[table]{xcolor}
\usepackage{listings}
\usepackage[colorlinks,bookmarksopen,bookmarksnumbered,citecolor=red,urlcolor=red]{hyperref}

\input{definitions} 

\usepackage{soul}

\usepackage{float}

\newcommand{\edited}[1]{#1}

\begin{document}

\ifnme
\runningheads{Wong, Kuhl, Darve}{A New Sparse Matrix Vector Multiplication GPU Algorithm}
\fi

\author{J.\ Wong,$^1$ \, E.\ Kuhl,$^1$ \, E.\ Darve$^{1,2}$
\ifnme
\corrauth
\else
\\[6pt]
$^1$Mechanical Engineering Department, \\
$^2$Institute for Computational and Mathematical Engineering,\\
Stanford University
\fi}
\title{A New Sparse Matrix Vector Multiplication GPU Algorithm Designed for Finite Element Problems}
\ifnme
\address{496 Lomita Mall, Durand Rm 226, Stanford, CA, 94305}
\corraddr{496 Lomita Mall, Durand Rm 226, Stanford, CA, 94305}
\fi

\ifnme
\begin{abstract}
\else
\maketitle

\section*{Summary}
\fi
Recently, graphics processors (GPUs) have been increasingly leveraged in a variety of scientific computing applications. However, architectural differences between CPUs and GPUs necessitate the development of algorithms that take advantage of GPU hardware. As sparse matrix vector multiplication (SPMV) operations are commonly used in finite element analysis, a new SPMV algorithm and several variations are developed for unstructured finite element meshes on GPUs. The effective bandwidth of current GPU algorithms and the newly proposed algorithms are measured and analyzed for 15 sparse matrices of varying sizes and varying sparsity structures. The effects of optimization and differences between the new GPU algorithm and its variants are then subsequently studied. Lastly, both new and current SPMV GPU algorithms are utilized in the GPU CG Solver in GPU finite element simulations of the heart. These results are then compared against parallel PETSc finite element implementation results. The effective bandwidth tests indicate that the new algorithms compare very favorably with current algorithms for a wide variety of sparse matrices and can yield very notable benefits. GPU finite element simulation results demonstrate the benefit of using GPUs for finite element analysis, and also show that the proposed algorithms can yield speedup factors up to 12-fold for real finite element applications.
\ifnme
\emph{\journalnamelc}.
\end{abstract}

\keywords{class file; \LaTeXe; \emph{\journalabb}}

\maketitle
\fi


\section{Introduction}
The recent use of graphics processors (GPUs) for scientific applications has opened up possibilities in achieving real-time simulations, and has been shown to provide remarkable speedup factors for many different applications. Scientific applications using GPUs range from computing the flow over hypersonic vehicles \cite{elsen08} to finite element simulations of virtual hearts \cite{haase10} and of viscoelastic properties of soft tissue \cite{taylor09} to medical applications~\cite{pratx11,joldes2010real,taylor2008high}. There is a developing body of literature concerning GPU implementation of finite element analysis codes.

Early papers investigating the use of GPUs for scientific computing focused on conjugate-gradient (CG) and multigrid solvers \cite{buatois09,bolz03,goddeke09}. Several studies in the area of finite-element analysis (FEM) have focused on the solution of the sparse linear system of equations $Ax=b$ resulting from a FEM discretization~\cite{Goddeke_FEMGPU, Goddeke_GPUMultigrid, Bolz_SparseMatrix}, mainly because the solution stage is often the most computationally intensive step.
Some assembly strategies for the GPU have been mentioned as well. Often, specific applications have allowed special approaches for FEM assembly. The methods described in~\cite{Bolz_SparseMatrix} for geometric flow on an unstructured mesh and in~\cite{Rodriguez_ClothGPU} for FEM cloth simulation derive relatively simple expressions for each non-zero in the system of equations. The relative simplicity and the inherent parallelism of computing each non-zero independently makes these approaches well suited for GPUs. The PhD thesis of G\"oddeke~\cite{goddeke2011fast} has an extensive discussion of the history behind GPU computing and the application of multigrid to FEM and its optimization for GPUs. Applications to the FEM package FEAST~\cite{Turek2010a} are discussed in~\cite{goddeke2011fast}.

Formulations have been considered in connection with discontinuous Galerkin methods. See~\cite{Klockner_DG_GPU} where the method is applied to Maxwell's equations. For high-order, continuous Galerkin methods, an assembly strategy is proposed in~\cite{Komatitsch_ElemAssembly,komatitsch2010high}. Other successful approaches, such as~\cite{Tejada_BodiesGPU} for deformable body simulation, took advantage of vertex and fragment shaders on the GPU to perform the assembly. New hardware and a more general computing environment appear to have made some of these techniques obsolete. Markall et al.~\cite{markall2013finite} compare FEM implementations on CPUs and GPUs. This paper includes a discussion of FEM along with spectral element methods and low-order discontinuous Galerkin methods. Recent investigations include~\cite{cecka11,cecka2010application,dziekonski12_1,dziekonski12_2}.

The difficulty in optimizing FEM calculations for GPUs has led some groups to develop special purpose languages and compilers to automate part of the process and separate the task of implementing novel numerical methods from the task of optimizing the computer implementation. \cite{markall2010towards} discusses  the Unified Form Language (UFL) for this purpose.

While GPUs are already used for many inherently parallel operations, such as material point O.D.E.\ solvers for finite elements \cite{lionetti2010, vigmond09} and for finite differences \cite{bartocci11}, total  performance gains for finite elements on GPUs can only be realized by efficiently combining all aspects of the calculation: element calculations, global assembly, and solver.

In this work, one of focuses is the sparse matrix solver that is used as part of FEM. However, it is generally not possible to simply ``port'' CPU algorithms to the GPU. Typically such implementations result in a GPU code that may be slower than the CPU code. Differences in GPU memory architecture, speed of individual GPU cores, and the communication between GPU cores, GPU memory, and host (CPU) memory can greatly impact the performance of GPU algorithms. Because of these differences, much effort has been spent in developing new algorithms designed for general purpose use.

The performance of many of these algorithms is dependent on the structure of the data. In particular, sparse matrices can be represented by a variety of formats and the computational performance may differ greatly depending on the format used~\cite{bell09}. There is a large body of literature on the investigation of various formats and their efficiency.

A sparse matrix contains mostly null entries. As a result it is advantageous to store only the non-zero entries along with their location in the matrix. The difficulty with such a storage is that it leads to a random access to memory, such that the performance is heavily limited by memory bandwidth and often very low. As a result several storage schemes attempt to make the memory access much more regular such that the effective memory bandwidth increases. This is crucial on GPUs where coalesced~\cite{nvidia_programming} memory access (i.e., accessing a 128-byte aligned segment in the device memory) is crucial for best performance. Bell et al.~\cite{bell2008efficient,bell09} was an early paper on SMPV and discusses all the key sparse matrix formats including the diagonal format (DIA); ELLPACK, in which all non-zero entries are stored in an $N \times K$ dense matrix ($N$ is the number of rows in the matrix and $K$ the maximum number of non-zeros per row); the coordinate format (COO, with a simple {\tt row}, {\tt col}, {\tt data} format); compressed sparse row format (CSR, this is a popular format on CPUs); and the hybrid format that combines ELLPACK with COO. At the end of the paper, the packet format (PKT) is discussed. In this format the matrix is decomposed into blocks with a high density of non-zero entries and a type of ``local'' CSR format is used to store entries associated with each block (offsets from the base index are stored using 16-bit integers).

Before exploring in more details various papers that followed the footsteps of Bell, we mention the CUSP library~\cite{Cusp} by Bell and Dalton which implements various matrix formats and iterative schemes. cuSPARSE~\cite{CUSPARSE} is a library released by NVIDIA which contains code for SPMV, SPMM (sparse matrix-matrix addition and multiplication), sparse triangular solve, a tri-diagonal solver, and incomplete factorization preconditioners.

Among the newer implementations, we start with a series of papers on variants of the ELLPACK format. ELLPACK-R~\cite{vazquez09} uses an ELLPACK format and each thread computes a single row. An additional array is used to store information about the true number of non-zero elements per row (not accounting for the ELLPACK padding), such that each thread is not required to do more floating-point operations than strictly necessary. This reduces the number of extra flops usually found in ELLPACK. Vazquez~\cite{vazquez10} revised his original algorithm and renamed it ELLR-T in 2010. In this work, he uses $T$ threads to compute a single row in SPMV. The Sliced ELLPACK~\cite{monakov10} format is similar to ELLPACK. However it is more flexibile and uses a storage scheme that varies the number of non-zero entries stored across CUDA warps. This amounts to an ELLPACK where $K$ is a function of the warp index or slice. The Sliced ELLPACK and ELLR-T (many threads per row) ideas are combined in the Sliced ELLR-T scheme of Dziekonski et al.~\cite{dziekonski11}.

A series of papers considered the block compressed sparse row format (BCSR). BCSR stores the non-zero entries as a sequence of fixed-size $r \times c$ dense blocks. BCSR uses one integer index of storage per block instead of one per non-zero as in CSR, reducing the index storage by $1/(rc)$. Moreover, fixed-sized blocks enable unrolling and register-level tiling of each block-multiply. The speedup achieved is unfortunately mitigated by fill-in of explicit zeros resulting from this storage. Vuduc et al.~\cite{vuduc05} mitigated these issues by splitting the matrix into a small set of matrices, each with its own $(r,c)$ and using a more flexible storage scheme called UBCSR. In~\cite{buatois09}, the BCSR scheme (called in that paper BCRS --- block compressed row storage) is used to implement a sparse general-purpose linear solver (i.e., Jacobi-preconditioned Conjugate Gradient). Monakov~\cite{monakov09} explored a variant with a hybrid BCSR/BCOO format. BCSR uses a CSR-like format to store blocks. The column coordinate for each block is stored, and the row coordinate is implicitly encoded by sorting blocks by rows and then storing the index of the first block in each row in a CSR-like format.  On the other hand, BCOO uses a blocked coordinate format where both coordinates of a block are stored. This provides additional flexibility in comparison to a pure BCSR format.

We mention a separate line of research by S.~Baxter~\cite{mgpu}. In the ModernGPU library, Baxter considers a SPMV as a sequence of two-steps. Step 1 is comprised of element-by-element multiplication of matrix entries by vector entries. This step is extremely parallelizable and runs near peak performance on GPUs. Step 2 is comprised of a segmented reduction (i.e., a large sequence of short reductions) to produce the final output vector. The focus is therefore on implementing an efficient segmented reduction. The resulting algorithm has high efficiency but is also very complex to implement and modify. The overall approach is very different compared to other SPMV papers. Nonetheless it has produced some of the fastest SPMV GPU codes currently available.

Our new algorithm follows in the footsteps of the ``padded jagged diagonals storage'' (pJDS) scheme~\cite{kreutzer12}. This is one of the most efficient storage formats available. It reduces the memory footprint by up to 70\%, and achieves 95\% to 130\% of the ELLPACK-R performance. A main drawback of ELLPACK-R and its sliced version is that when rows are highly inhomogeneous (the number of non-zeros varies significantly), memory and floating-point operations are wasted and this overhead may become notable. In contrast, pJDS sorts all the rows from longest to shortest. The rows are then grouped into slices (as in sliced ELLPACK-R), and then packed with zeros up to the longest row. Because of the sorting, the extra padding is minimal and most floating-point operations become more efficient. Here is the pseudo-code for this algorithm:
\begin{figure}[htbp]
\begin{lstlisting}[language=C++,
caption=Pseudo-code for the pJDS scheme,
label=jJDS]
for (i=0; i < N; ++i)
  for (j=0; j < rowmax[i]; ++j){
    col_offset = col_start[j];
    c[i] += val[col_offset + i] * rhs[ col_idx[col_offset + i] ];
  }
}
\end{lstlisting}
\caption{N is the number of rows in the matrix, while the array val holds the elements to the matrix. rhs is the vector to be multiplied in SPMV and col\_idx are the column indices pertaining to the elements of val. rowmax designates the length of a particular row. col\_offset denotes the start of a particular column of the matrix val.}
\end{figure}

The permutation on the rows of the matrix means that entries in the vector should be permuted as well. There is a computational cost associated with that. However in the context of iterative methods, this overhead is mitigated. The approach in this paper has a similar drawback.

In developing a GPU-based finite element code for real-time heart simulations, we developed a variant of pJDS. We found that it works very well for finite element unstructured grids as well as a variety of other sparse matrices as found in the benchmark suite of Williams et al.~\cite{Williams2009178,matrix_bench}. Our algorithms take a novel approach by systematically addressing issues with current GPU algorithms related to row length irregularity, padding efficiency of matrices, and data localization for GPUs. While other algorithms have similar insights in gaining additional performance for finite element SPMV operations by grouping rows with the same row structure \cite{vuduc05,kreutzer12}, encoding sparse matrices in blocks \cite{buatois09,monakov10,dziekonski11}, or utilizing data coalescing repacking algorithms \cite{vazquez10,mgpu}, our algorithms were designed to reduce GPU data locality issues, increase memory throughput, and perform well on relatively unstructured meshes with a substantial degree of irregularity in row lengths distributions.

Compared to the pJDS scheme (row sorting) discussed above, our algorithm differs in a few ways. We use a warp optimized storage. As in pJDS, rows are grouped together into sets (slice) such that a single warp processes a given set. Inside a set, the number of non-zero entries is taken to be constant and is set to be equal to the maximum exact number of non-zeros in the row set. Entries for a given set are stored contiguously in memory such that the memory access by threads is perfectly coalesced. A varying number of threads is assigned to a given row. This allows optimizing situations in which few rows are significantly longer than others. In that case, a large number of threads (or all 32 threads in a warp) can be assigned to that row. As a result, per warp, we basically only need to know 3 integers for memory access: the starting address in the matrix $A$, the number of rows processed by the warp, and the number of non-zero entries per thread. Because of the permutation on the rows of the matrix (as in pJDS), the sequence of column indices inside a given row is typically ``random''. We sorted them in order to improve cache performance when accessing the right-hand side vector $x$. In some sense our algorithm combines insights from ELLPACK-R (varying the number of non-zeros $K$), sliced format (reduced storage), ELLR-T (many threads per row; in our case the number of threads per row is not even constant and is adjusted for load-balancing), and pJDS (sorting of rows). We called our new scheme ELL-WARP.

\edited{During the review process, it came to our attention that an algorithm called SELL-$C$-$\sigma$ was proposed in a recent paper submitted for publication and available via ArXiv \cite{Kreutzer2013}. Their publication compares Intel, Intel Phi, and Nvidia architectures and they examine different metrics with which to evaluate their SELL-$C$-$\sigma$ algorithm. We developed our algorithms separately and concurrently, and our initial algorithm K1 is equivalent to SELL-$C$-$\sigma$. Their paper also highlights difficulties in addressing particular types of matrices which we have addressed fairly well in our K2 algorithm. This paper concentrates more on examining the GPU aspects of implementation, the important details regarding the necessary reordering costs of these classes of algorithms, and also proposes some techniques of ameliorating those issues. Lastly, while this paper confirms some of the observations of the aforementioned paper, we also provide further insight into other important aspects of SPMV computational performance differences by carefully modifying our proposed kernels.}

We also note that within the context of the finite element method, our novel SPMV routines can be leveraged both for the global assembly operation and within the solver to gain additional computational performance.

In this paper, we first describe the computational heart simulation problem within the context of a traditional finite element framework. We then highlight computational costs for both the assembler and solver, and highlight differences between CPU and GPU implementations. A survey of different GPU SPMV algorithms then follows. A new set of SPMV algorithms is introduced, and a variety of SPMV algorithms are benchmarked against our novel algorithms. Different aspects of optimizations used in the development of our algorithms are investigated. We discuss the utility of our SPMV algorithms in the contexts of finite elements and general computational settings. Lastly, we end with concluding remarks where we highlight possible future improvements to our algorithms, and also discuss possible improvements in leveraging GPUs for finite element simulations.

\section{Description of the Physical Problem}
Our original aim was to utilize GPUs for achieving real-time electrophysiological simulations of the heart. The electrical propagation of voltage within the heart is described by the following two-variable phenomenological governing differential equations for nonlinear mono-domain excitable tissues.
\begin{gather}
\dot{\phi} = \divg \vec{q}(\phi) + f^{\phi}(\phi,r)
\label{eqn:mono-domain1} \\
\dot{r} = f^{r}(\phi,r)
\label{eqn:mono-domain2}
\end{gather}
The transmembrane voltage, $\phi$, is the voltage difference across the cell membrane, while the phenomenological variable, $r$, describes the recovery behavior of cardiac tissue. To account for conduction throughout the tissue, a phenomenological potential flux term $\divg \vec{q}$ is added.
\begin{equation}
\vec{q} = \ten{D} \cdot \nabla \phi
\end{equation}

A phenomenological diffusion tensor $\ten{D} = d_{\scas{iso}} \ten{I} + d_{\scas{ani}} \vec{n} \otimes \vec{n}$ allows for cell-to-cell electrical coupling across cellular gap junctions. $d_{\scas{iso}}$ and $d_{\scas{ani}}$ are the respective isotropic and anisotropic conduction terms and $\vec{n}$ is the preferred direction of anisotropic conduction.

In this paper, we will use the classical Aliev-Panfilov model \cite{aliev96} for convenience to evaluate the effectiveness of our GPU algorithmns. The source terms for the Aliev-Panfilov model are
\begin{align}
f^{\phi} &= c \, \phi \, [\phi - \alpha][1-\phi] - r \, \phi \\
f^r
&= \left[ \gamma + \frac{\mu_1 \, r}{\mu_2 + \phi} \right]
\, [-r -c \, \phi \, [\phi-b-1]]
\label{eqn:ap}
\end{align}
While the mono-domain equations (\ref{eqn:mono-domain1}) and (\ref{eqn:mono-domain2}) can be solved simultaneously, a global-local internal variable splitting approach \cite{goktepe09,wong11} is taken because it yields a global symmetric tangent matrix for the global degree of freedom, $\phi$, such that a fast iterative GPU solver can be effectively employed. \edited{In our splitting approach, we take the transmembrane voltage, $\phi$, to be our global variable and $r$ to be our local variable. $r$ is local in the sense that the value of $r$ at a given point in time is not directly affect by the current values at the neighboring points. $r$ at a given point is affected by neighboring points through coupling through $\phi$.}

An implicit semi-discretization approach is taken where backward Euler time integration is used and the mesh is discretized spatially using Lagrangian shape functions. The heart mesh is discretized in space using isoparametric tetrahedral elements in this paper using equations (\ref{space01}) $N$ are tetrahedral shape functions, while $\delta \phi_{i}$ and $\phi_{j}$ are the respective test function local nodal values and local nodal solution values. ${\cal{B}}^e$ is the domain of the element and $n_{\rm{en}}$ is the number of nodes within a particular element.
\beq
 \delta  \phi^h|_{{\cal{B}}^e}
=\!\sum_{i=1}^{n_{\rm{en}}} N^i \; \delta \phi_{i}
 \qquad
 \phi^h|_{{\cal{B}}^e}
=\!\sum_{j=1}^{n_{\rm{en}}} N^j \; \phi_{j}
\label{space01}
\eeq

\edited{After the initial spacial discretization has been performed, initial values are set for the global variables ,$\phi_J$, at the finite element nodes, while the internal variable $r$ is initialized at each integration point within each element. As the coupled nonlinear problem is stiff, backward Euler implicit time integration is used where $\phi_n$ denotes the fully converged solution at the previous timestep.
\[
\dot{\phi} \approx \mbox{$\frac{1}{\Delta t}$} [\,\phi - \phi_n\,]
\]

For a given timestep, $n+1$, Newton's method is applied over the finite element nodal values, $\phi_J$, such that the global residual, $\scas{R}^{\phi}$ is fulfilled. In order to ensure that the residual for the internal variable, $\scas{R}^{r}$ is also fulfilled over the whole domain, we apply Newton's method at every integration point within every element and update $r$ such that $\scas{R}^{r}$ is fulfilled at every point. Therefore,} the nonlinear rate equation (\ref{eqn:mono-domain2}) is first solved locally \edited{using equation \ref{eqn:localres}}. The tangent matrix $\sca{K}^{r}$ and residual $\sca{R}^r$ are constructed at each integration point and used to iteratively determine the converged nonlinear solution \edited{for the given timestep , $n+1$.}
\beq
\sca{R}^{r} = \dot{r} - f^{r}(\phi,r)
  \qquad
\sca{K}^{r} = \sca{d}_{r}\sca{R}^{r} = \frac{d\sca{R}^{r}}{dr}
\label{eqn:localres}
\eeq

\edited{With $r$ updated based on $\phi_J$, the current Newton iteration value of the transmembrane potential at the current timestep at each finite element node, the terms for the global residual, $R^{\phi}_{I}$ can be calculated using equations \ref{eqn:globalres}.} The tangent term $\sca{d}_{\phi}f^{\phi}(\phi,r)$ \edited{ necessary for computing $\scas{R}^{\phi\,\scas{e}}_I$, the global residual contributions from each element, can be computed in a straightforward manner and the tangent term is used to obtain the completed global element tangent matrix ${\sf{K}}_{IJ}^{\phi\,{\scas{e}}}$  for each element. The details for this procedure can be found here \cite{wong11}.} The global element residuals  ${\sf{R}}_I^{\phi\,{\scas{e}}}$ are then calculated and together with the element tangent matrices are assembled to form the global tangent matrix ${\sf{K}}_{IJ}^{\phi}$ and global residual vector ${\sf{R}}_I^{\phi}$ \edited{, as is done normally in nonlinear finite elements.} 
\beq
{\sf{R}}_I^{\phi\,{\scas{e}}} = \dot{\phi}^{\scas{e}} - \divg \vec{q}(\phi^{\scas{e}}) - f^{\phi}(\phi^{\scas{e}},r^{\scas{e}}) 
\qquad
{\sf{K}}_{IJ}^{\phi\,{\scas{e}}} = \sca{d}_{\phi_J^{\scas{e}}} {\sf{R}}_I^{\phi \,{\scas{e}}} = \frac{d{\sf{R}}_I^{\phi\,{\scas{e}}}}{d{\phi_J^{\scas{e}}}}
\label{eqn:globalres}
\eeq
 With the global tangent matrix and residual, the global finite element problem is solved iteratively using Newton's method over the domain.  Both the local and global degrees of freedom are updated at each global Newton-Raphson iteration to ensure the consistency of the solution. This process is repeated until convergence is reached at each timestep.
The scheme is summarized in Table~\ref{table_algorithm}.
\begin{table}[htbp]
\renewcommand{\arraystretch}{1.2}
\begin{center}
\caption{An algorithmic treatment of the transmembrane voltage in excitable cardiac tissue based on finite element discretization in space and implicit finite difference discretization in time embedded in two nested Newton-Raphson iterations. The electrical unknown, the membrane potential $\phi$, is introduced globally on the node point level whereas the phenomenological recovery variable, 
$r$, is introduced locally on the integration point level. Element assembly is highlighted in blue, and the solution update is highlighted in red.}
\label{table_algorithm}

\vspace*{0.4cm}
\resizebox{\textwidth}{!} {
\begin{tabular*}{16.05cm}
{@{\extracolsep{\fill}}|p{15.65cm}|} \hline 
initialize nodal degrees of freedom $\phi_J$ \\
initialize internal variable 
$r$\\[2.pt] \hline \hline
   global Newton iteration\\ [1.pt]
      \hspace*{0.29cm}\begin{tabular*}{15.55cm}
      {@{\extracolsep{\fill}}|p{15.15cm}|} \hline
      \rowcolor[rgb]{.5,.5,.8} loop over all elements \\ [1.pt]
      \rowcolor[rgb]{.5,.5,.8} \hspace*{0.29cm}\begin{tabular*}{15.05cm}
      {@{\extracolsep{\fill}}|p{14.65cm}|} \hline 
       \rowcolor[rgb]{.5,.5,.8} loop over all integration points \\ [1.pt]

      \rowcolor[rgb]{.5,.5,.8} \hspace*{0.29cm}\begin{tabular*}{14.55cm}
      {@{\extracolsep{\fill}}|p{14.15cm}|} \hline 
     \rowcolor[rgb]{.5,.5,.8} local Newton iteration \\ [1.pt]
      \rowcolor[rgb]{.5,.5,.8}\hspace*{0.29cm}\begin{tabular*}{14.05cm}
      {@{\extracolsep{\fill}}|p{13.65cm}|} \hline 
      \rowcolor[rgb]{.5,.5,.8} calculate local recovery variable residual,
      ${\sf{R}}^{r} = [r-r^{n}]/\Delta t - f^{r}\!$ and local tangent matrix 
      $[{\sf{K}}^{r}] \!
      = \! \sca{d}_{r} {\sf{R}}^{r}$
      \\
      \rowcolor[rgb]{.5,.5,.8} update the recovery variable $r \leftarrow r
      - [{\sf{K}}^{r}]^{-1} {\sf{R}}^{r}$ 
      \\ [2.pt] \hline
      \end{tabular*} 
      \\ \rowcolor[rgb]{.5,.5,.8} calculate source term $f^{\phi}(\phi,r)$
      and its linearization $\sca{d}_{\phi}f^{\phi}$
      \\ [2.pt] \hline
      \end{tabular*} 
      calculate element residuals ${\sf{R}}_I^{\phi\,{\scas{e}}}$
      and element matrices ${\sf{K}}_{IJ}^{\phi\,{\scas{e}}}
      =\sca{d}_{\phi_J^{\scas{e}}} {\sf{R}}_I^{\phi \,{\scas{e}}}$ 
      \\ [2.pt]\hline
      \end{tabular*} 
      calculate global residual ${\sf{R}}_I^{\phi}$
      and global iteration matrix ${\sf{K}}_{IJ}^{\phi}
                           =\sca{d}_{\phi_J} {\sf{R}}_I^{\phi}$  
      \\ [2.pt] \hline
      \end{tabular*} 
   \\
    \rowcolor[rgb]{.8,.5,.5} 
   update membrane potential $\phi_J \leftarrow \phi_J 
   - {\sf{K}}_{IJ}^{\phi -1} \, {\sf{R}}_I^{\phi}$ 
   \\ [2.pt] \hline 
\end{tabular*} 
}
\end{center}
\end{table}
%

Implicit time integration can be further leveraged by using adaptive time-stepping schemes to further reduce the computation time. Lastly reduced-order integration is used, whereby a linear tetrahedral element will only have one integration point. All implementations developed in this study are scientifically accurate and run at double precision.

\subsection{Computational Cost of the Finite Element Method}
\begin{figure}
\begin{center}
\includegraphics[width=\textwidth]{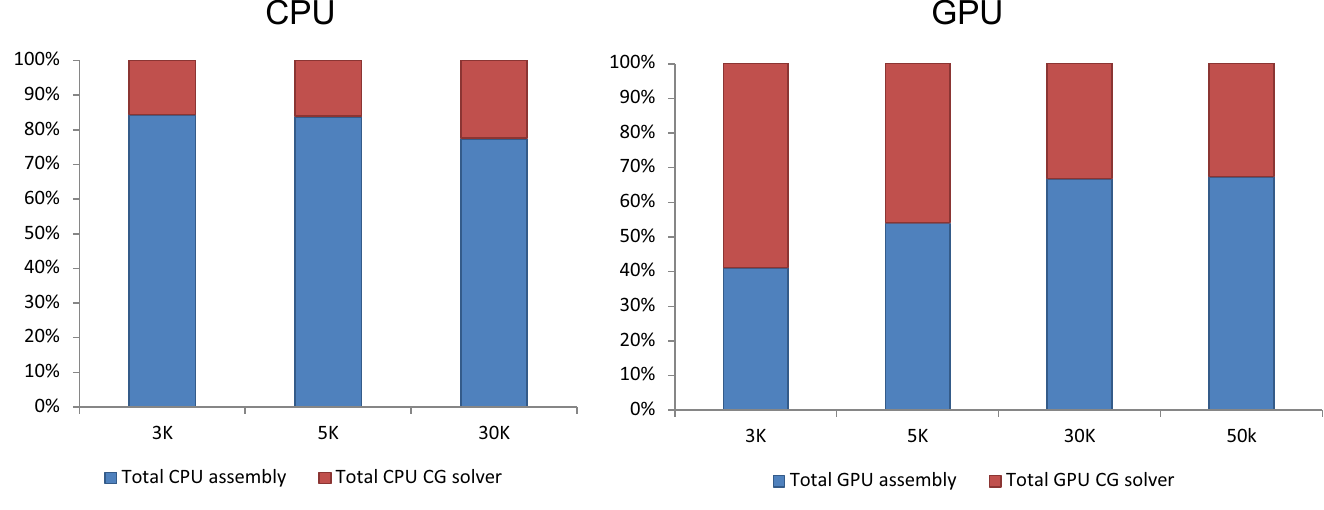}	
\end{center}
\caption{Percentages of computation time of finite element assembly (red) and the CG solver (blue) of the heart problem compared to the total simulation time are shown for both the CPU (left) and GPU (right). Results are shown for the same finite element heart geometry at different refinements: 3K, 5K, 30K, and 50K nodes. A Jacobi preconditioner was used for both the CPU and GPU solver. The PETSc solver using a single core failed to converge for the 50K mesh when the simple Jacobi preconditioner was used.}
\label{fig:CPUGPUcomparison}
\end{figure}

While the biophysical problem may initially seem rather specific, the problem has been structured in the standard non-linear solid mechanics finite element framework. For example, such local-global splitting schemes are commonly used in the field of plasticity \cite{simo88}. A comparison between the computational cost of our heart simulation for various refinements of the same mesh are shown in Figure \ref{fig:CPUGPUcomparison} for a single CPU finite element method implementation of the electrophysiological problem using a standard scientific code, PETSc \cite{PETSC}, and for our custom single GPU implementation.A Jacobi preconditioner was used for both the CPU and GPU solver. Four different mesh refinements were created having roughly 3,000 nodes, 5,000 nodes, 30,000 nodes, and 50,000 nodes. These meshes have 11k, 16k, 123k and 218k elements respectively.
Our GPU implementation utilizes sparse matrix vector (SPMV) multiplication operations to assemble the global tangent and residual quantities, and SPMV is also employed in our standard Jacobi preconditioned conjugate gradient solver. The use of SPMV within the context of CG solvers is prevalent \cite{ buatois09, bolz03, goddeke09, vuduc05, monakov09}; however the benefit in using SPMV for global assembly on GPUs is not particularly obvious and will be explained in more detail in the following sections.

\begin{figure}[htbp]
\begin{center}
\includegraphics[width=\textwidth]{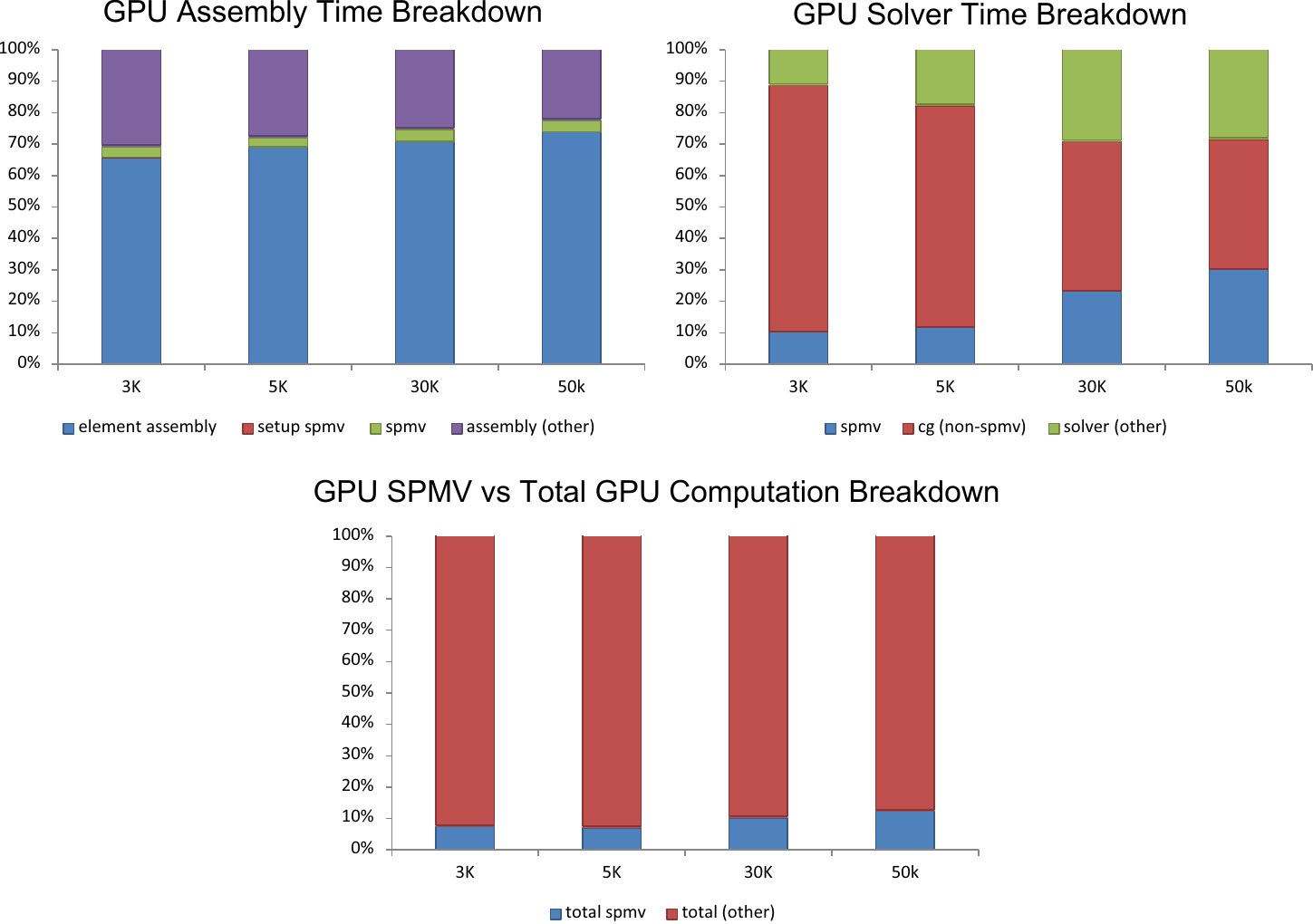}
\end{center}
\caption{The proportion of time spent on different finite element assembly tasks within our GPU implementation for different refinements is shown (top left). SPMV is used both during the assembly and during the CG solver. Element assembly was performed on a thread level where each thread assembles an element and is shown in blue. The time spent preparing for the SPMV operation is shown in red, while the actual SPMV operation is denoted in green. The time spent initializing, organizing, and copying to and from GPU memory is denoted in purple.  Proportions of SPMV with relation to other tasks for the CG solver are shown (top right). The time spent performing the SPMV operation (blue), the time spent performing the CG algorithm excluding time spent performing the SPMV (red), and initialization and GPU memory transfer (green) are indicated for each mesh refinement. The total proportional time spent on SPMV for both assembly and solver portions (blue) in comparison to the rest of the GPU finite element method (red) is shown (bottom). CUSPARSE \cite{CUSPARSE} was used to carry out all SPMV operations in these results.}
\label{fig:GPUspmv}
\end{figure}

For our GPU heart simulations the portion of time spent on SPMV during assembly is relatively small compared to the total assembly time and increases towards 33\% within the CG solver as the level of refinement increases (Figure \ref{fig:GPUspmv}). Together, the total time spent in our GPU implementation shows that up to 10\% of the computation time is spent purely on SPMV calculation. However, if other parts of the code are further optimized and accelerated, it is possible that the SPMV calculation could become a considerable bottleneck. Thus, while general finite element problems may exhibit different solver-to-assembly ratios, improvements to SPMV operations can be beneficial to finite element frameworks. In the following sections, we will investigate various improvements and novel modifications to current SPMV algorithms, and highlight their use within the finite element method and in other general SPMV applications.

\section{GPU Architectural Organization}
As GPUs are architecturally organized differently than traditional CPUs, a brief summary is given before we examine current GPU algorithms. These architectural differences are important in developing a substantially improved SPMV algorithm.  Since we are using an Nvidia GPU, we will describe the general organization of an Nvidia GPU. The overall work-execution model on the GPU is composed of individual threads that each execute a specified kernel routine concurrently with respect to other threads (Single-Instruction Multiple-Thread = SIMT). Groups of 32 threads, called warps, are executed synchronously. Furthermore, up to 32 warps can be grouped into a block (CUDA Compute Capability 2.0). Threads from a block have access to shared memory within that block and can also be forced to synchronize with other threads within that block.

On the GPU, there are basically four types of memory available: global, shared, local, and register memory. In general, memory that is available to more threads is slower than memory that is shared within a smaller group of threads. For example, global memory can be accessed by all threads on a GPU, but it is also the slowest form of memory available. Shared memory is shared within a block. It is always faster than global memory. However memory bank conflicts may occur when threads are accessing the same shared memory banks simultaneously. Local memory is global memory reserved for threads where stores are kept in the L1 cache. Generally its use is specified by the compiler and it is used when a thread has fully utilized the maximum amount of memory registers available. Registers are the fastest form of memory, and only 63 registers are available for a given thread for the GPU used in this work. Different memory caches are used to facilitate memory accesses to and from the different types of memory to the individual threads. In this paper, the L1 and L2 caches are implicitly used and use of the texture cache, which is a special spatial locality-based global memory cache, is also investigated.

\section{Description of Current GPU Methods}
In order to achieve real-time finite element heart simulations, ideally both the finite element tangent matrix assembly and finite element solver will perform a majority of the computations on the GPU to minimize CPU-to-GPU memory transfer overhead. Current GPU approaches and methods are summarized below.

For the finite element assembly of cardiac simulations, efforts have been made at leveraging GPUs to speed up the evaluation of the local O.D.E.\ problem (\ref{eqn:mono-domain2}) \cite{lionetti2010, vigmond09}. Since we have chosen a simple phenomenological model, these efforts are not directly applicable to this study. However, several different finite element assembly approaches have been recently~\cite{markall2013finite, cecka11, Huthwaite2014, Plaszewski2010}. In Cecka et al.\ it is shown that assembly by non-zeros of the finite element tangent matrix is a better approach compared to assembly by elements, which is the traditional serial finite element approach. Furthermore, if we restructure the assembly such that shared memory is used instead of global memory, further speedups are possible. However, this improvement requires a substantial rewrite of the finite element method and is dependent on the particular type of element. The method of coloring, where elements are specified in a way such that threads can safely accumulate entries in the global matrix without potential race conditions, can also be used, but multiple passes are typically required and memory traffic may increase. These consequences may affect the overall performance, however the number of flops is often nearly optimal. Also of note, in Markall et al. 2013, a method for assembly the residual vector is proposed as the best solution for their problem; however they propose a different scheme for the global matrix assembly. Much work has also been done on finite element CG solvers and multi-grid solvers on GPUs \cite{haase10, buatois09, goddeke09, vuduc05}. Lastly, both the finite element assembly on the GPU and conjugate gradient method utilize forms of sparse matrix vector product multiplications.

In this paper, we take a hybrid approach where element quantities are first obtained in parallel. However instead of directly assembling the global tangent matrix directly from each finite element's stiffness matrix, these element quantities, ${\sf{K}}_{IJ}^{\phi\,{\scas{e}}}$, are organized into an SPMV problem where each row corresponds to the assembly of a non-zero element in the global matrix, ${\sf{K}}_{IJ}^{\phi}$. This approach attempts to avoid race conditions and retain efficiency while avoiding large deviations from standard finite element codes. In our approach, the finite element assembly procedure can be reorganized into two separate SPMV operations to build the tangent matrix, $\sf{K}$, and residual vector, $\sf{R}$, for non-linear iterative finite element simulations. While it is possible to combine the assembly into one operation where the memory associated with $\sf{K}$ and $\sf{R}$ is adjacent, for the sake of clarity in this paper we will treat them as separate SPMV operations.
\begin{gather*}
\ten{K}^{\scas{elem}} =
\begin{bmatrix}
\sf{K}^{\phi\,\scas{1}}_{\scas{1,1}}&\sf{K}^{\phi\,\scas{2}}_{\scas{1,1}} & \sf{K}^{\phi\,\scas{3}}_{\scas{1,1}} & \sf{K}^{\phi\,\scas{4}}_{\scas{1,1}} & \hdots \\
\sf{K}^{\phi\,\scas{1}}_{\scas{1,2}}& \sf{K}^{\phi\,\scas{4}}_{\scas{1,2}} & 0 & \hdots \\
\vdots \\
\sf{K}^{\phi\,\scas{7}}_{\scas{n_{nodes},n_{nodes}}}& 0 & \hdots
\end{bmatrix}
,\qquad
\ten{K}^{\scas{elem}} \cdot \ten{\bar{1}}
= \ten{\sf{K}}^{\phi}_{\scas{flat}}
\end{gather*}
In our problem, $\vec{K}^{\scas{elem}}$ is the temporary sparse tangent matrix used to assemble the global tangent matrix. Each row corresponds to a particular entry, ${\sf{K}}^{\phi}_{\scas{IJ}}$. Therefore $\vec{K}^{\scas{elem}}$ has $\scas{nnz_{K}}$ rows, where  $\scas{nnz_{K}}$ is the number of nonzero global elements in $\ten{\sf{K}^{\phi}}$. Entries within a row are simply the different contributions from the various element stiffness matrices $\sf{K}^{\phi\,\scas{e}}_{\scas{IJ}}$ associated with the particular tangent matrix location. By summing up each row the global matrix entries, ${\sf{K}}_{IJ}^{\phi}$ can be accumulated without any race conditions in parallel. This, in fact, can be described as a trivial SPMV operation (shown above). $\ten{\bar{1}}$ is a vector of 1's of length $\max(n^{\scas{i}}_{\scas{conn}})$, where $n^{\scas{i}}_{\scas{conn}}$ is the number of elements that share a node $\scas{i}$. The resulting global matrix $\sf{K}^{\phi}$ can be flattened into a vector, $\sf{K}^{\phi}_{\scas{flat}}$, that has $\scas{nnz_{K}}$ entries. In the same way, the residual vector entries, $\vec{R}^{\scas{elem}}$, can be assembled using the aforementioned SPMV organization used for assembling $\sf{R}^{\phi}$. $\vec{R}^{\scas{elem}}$ is a matrix of size $n_{\scas{nodes}} \times \max(n^{\scas{i}}_{\scas{conn}})$. We note that it is trivial to modify a SPMV algorithm such that only summations are performed without needing to multiply each elemental quantity by 1.

The CG solver is then used to solve the canonical problem $\ten{\sf{K}}^{\phi} \, \Delta\vec{x} = \vec{\sf{R}}^{\phi}$, where $\Delta\vec{x}$ is the solution update for the Newton-Raphson method. At least one SPMV is performed during each CG iteration step. Therefore, by improving SPMV algorithms on GPUs, we can improve different parts of the finite element method and also apply these algorithms to general SPMV linear algebra problems.

While there are many variations of sparse matrix vector product GPU algorithms \cite{monakov09, buatois09, vuduc05}, the majority can be summarized by the available implementations from several commonly available sparse matrix vector libraries: CUSPARSE \cite{CUSPARSE}, CUSP-library \cite{Cusp}, and ModernGPU \cite{mgpu}. In \cite{bell09}, COO, CSR, ELL, and HYB formats are examined and analyzed. A review of the different formats and algorithms is highlighted below. In this section, we will look at the classical SPMV problem $\ten{A} \, \vec{x} = \vec{y}$, where $\vec{x}$ is given and the objective is to calculate the vector $\vec{y}$.

\subsection*{COO}
The coordinate list (COO) sparse matrix format is composed of three array lists of row indices, column indices, and the corresponding list of matrix values. In the CUSP COO implementation, row indices are sorted in order. Each warp processes a section of the matrix and works over 32 non-zero values at a time. Each thread within the warp performs a multiplication between the thread's value and the corresponding value from the vector. The different row segments are summed using segmented reduction, which is a method of efficiently distributing the reduction over a warp even if the warp is computing the sum for different rows. During the initial stage, the product is stored in shared memory and intra-warp segmented reduction is performed on the element products. The first thread within each warp determines whether to include results from a previous warp iteration in its row sum, and if not, updates the solution vector. Intra-block segmented reduction is then used to properly accumulate rows that span multiple warps.

COO SPMV algorithms generally suffer from poor memory to computation ratios as row and column indices must be retrieved for each computation. Row indices are also required for row sum reduction, and additional explicit intra-block thread synchronization may be required depending if rows span multiple warps.

\subsection*{CSR}
The compressed sparse row (CSR) matrix storage format is composed of three array lists of values, corresponding column indices, and corresponding row offsets that index the beginning of each row in the arrays of values. In the CUSP CSR vector implementation, a warp is assigned to each row in the matrix. A warp processes a continuous section of values and corresponding column indices in a coalesced manner. Intra-warp reduction is performed and sums are accumulated in the solution vector.

The main issue with CSR algorithms is that while the storage is space efficient and the memory access pattern is contiguous, memory accesses are not aligned. The CUSP implementation will also suffer when the size of a row is less than the warp size. However, unlike an alternative CSR algorithm where each thread is assigned to a row, the access pattern in CUSP is coalesced and contiguous.

\subsection*{ELL and ELL variants}
In the ELLPACK (ELL) format, the matrix is again described by two array lists of values and column indices. The maximum size of the longest row in the matrix is allocated for each row in the ELL format. Non-zero values are ordered contiguously, and the remaining values are typically zero-padded. Column indices are arranged and padded in a similar manner. Both values and columns are stored in column-major order. The GPU kernel is fairly simple. Each thread is responsible for a row, and matrix vector products are summed in a coalesced manner.

The ELL format performs poorly when the row size varies from row to row resulting in poor work distribution within each warp. The format also suffers from issues with excess padding when the longest and shortest row differ greatly in length, and when the average row length is significantly shorter than the longest row.

There are several variations of the ELL format which attempt to fix these issues. One is the ELL-R format \cite{vazquez09} where an extra vector is provided that designates the number of non-zeros each row is responsible for. Another is Sliced ELLPACK \cite{monakov10} where the matrix is sliced into groups of consecutive rows and then stored in the ELLPACK format. The ELLR-T \cite{vazquez10} and sliced ELLR-T formats \cite{dziekonski11} use multiple threads per row. Lastly, another variant is the padded jagged diagonals storage (pJDS) format \cite{kreutzer12} which is similar to ELL-R. In this format, the rows are first sorted from the longest to shortest row, and then sliced and padded in a manner similar to the ELL-R format.

\subsection*{HYB}
The Hybrid format is similar to the ELL format. To reduce the amount of zero-padding due to row-to-row size differences, a certain number of values, determined empirically, is stored in ELL, and the remaining entries are stored in COO format. The CUSP implementation is simply a combination of the ELL kernel and the COO kernel.

Due to the more efficient storage scheme, the hybrid format requires launching of two kernels in order to perform the sum. While the values of the matrix are contiguous in each kernel, the access pattern for the vector will generally not be so.

\subsection*{ModernGPU}
ModernGPU takes a different approach with its sparse matrix vector implementation called mgpusparse (MGPU). The algorithm is somewhat complicated but is essentially constructed to address several issues with the algorithms above. It is a combination of a series of parallel and segmented scans. Data is partitioned into the list of values, list of corresponding column indices, and several lists are used to keep track of the ends of rows and row segments. Data is also arranged in a coalesced way similar to column-major storage, but for fixed-sized chunks of matrix values. Two kernels are utilized. In the first kernel, each thread within a warp is responsible for calculating \texttt{valuesPerThread} partial row sum entries. A serial scan is performed by each thread and the sums of different row segments are stored in shared memory. A number of flags are associated with each thread, such that each thread can determine where to store the partial sum for a given row segment in shared memory. An intra-warp parallel scan is then performed to reduce rows within each chunk quickly. Lastly an intra-block segmented reduction is performed to reduce rows that span multiple blocks.

MGPU also takes advantage of index representation compression and performs other compression tricks to increase efficiency. The algorithm generally performs well in comparison to the algorithms above. The main drawback, however, is the complexity of the algorithm and the amount of book-keeping that is necessary. The main benefits are that the workload for threads within a block are very well distributed and very few threads are idle compared to the rest of the threads within each block.

\section{Development of Novel GPU Algorithms}
There are several other variations of the previously mentioned algorithms. Most variations concentrate on increasing padding efficiency, attaining less warp-divergence, and increasing the workload for a given block of threads \cite{buatois09,vuduc05}. However, there are several key insights that have been made in ModernGPU, ELLPACK, and CSR vector implementations that may yield a better SPMV algorithm for finite elements. In ELL and MGPU algorithms, ``column-major" coalesced data interleaving is leveraged to allow threads within a warp to work on different rows and row-segments. The interleaving results in an efficient coalesced memory access pattern. In MGPU and CSR vector, intra-warp reductions are utilized to avoid unnecessary thread synchronization, which allow warps and blocks to execute without delays due to synchronization.

The variations in row length in ELL are acquiesced in the HYB matrix implementation, and are accounted for by segmented scan flags in MGPU. Both algorithms try to address the padding inefficiency in the simple ELL algorithm. In \cite{bell09}, it is noted that the impact of padding on well-behaved structured meshes should not be substantial as the row size should not vary excessively. In unstructured finite element meshes of solids, this appears to be partially true. Since the node connectivity of the mesh directly translates into the finite element tangent matrix and element quality is usually controlled by meshing algorithms, the number of non-zeros per row is indirectly controlled by mesh quality algorithms. Therefore, the row lengths in a given unstructured mesh are usually constrained; however, there is no guarantee that there is row-to-row uniformity within the matrix for a given unstructured finite element mesh. In the case of assembly, the $\vec{x}$ vector is simply a vector of ones, and one can simply avoiding the multiplication of unity to the corresponding matrix value altogether.
%
\begin{figure}[htbp]
\begin{center}
\includegraphics[width=0.7\textwidth]{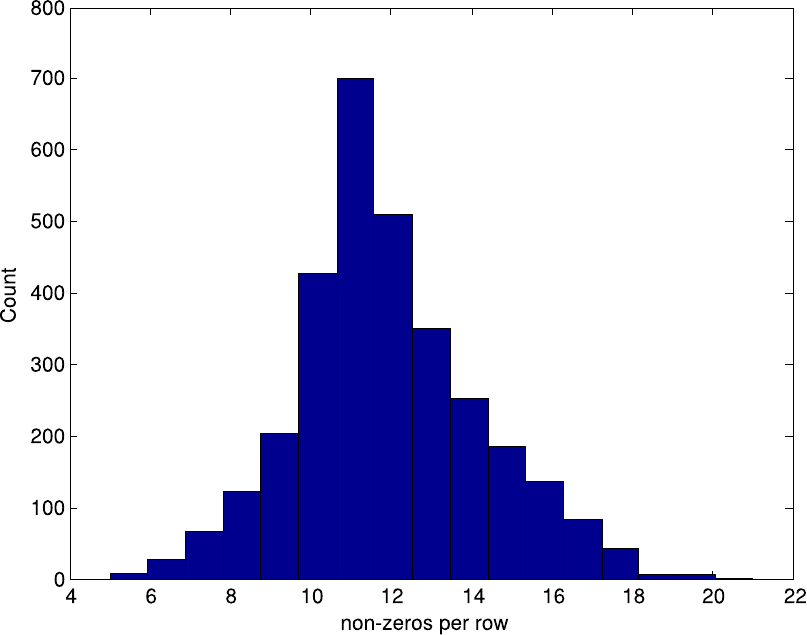}
\end{center}
\caption{A histogram of the non-zeros per row in a small 3129 row stiffness matrix for a patient-specific simulation of the heart (Heart 3K).}
\label{fig:nzperrow}
\end{figure}
From Figure \ref{fig:nzperrow}, it is clear that while row sizes only vary from 5 to 21 non-zeros per row, the actual distribution of non-zeros per row varies quite a bit from the median and mean row lengths. This distribution poses problems for ELL because the row size varies from row-to-row and therefore the throughput may not be optimal. The distribution also does not yield optimal throughput for CUSP CSR since row sizes are smaller than the size of a warp.

Since MGPU outperforms the simple kernel implementations of ELL/HYB, this seems to indicate that there may be inefficiencies related to padding which are not an issue for the more complicated equal work distributed MGPU implementation. Therefore, we propose the following set of simple kernel algorithms that are directly motivated from the insights listed above.During the preparation of this manuscript, it came to our attention that other ELL variations --- sliced ELLR-T and pJDS --- share many features with our proposed kernels. While we came up with our algorithms independently, we have made similar key observations. We aim to show how our algorithms are a generalized superset of the ELL variations.

\subsection{ELL-WARP}
\begin{figure}[htbp]
\centering
\includegraphics[width=0.7\textwidth]{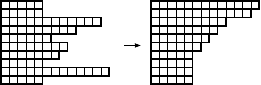}
\caption{Sort rows of the matrix from longest to shortest.}
\label{fig:sort}
\end{figure}

\begin{figure}[htbp]
\centering
\includegraphics[width=0.7\textwidth]{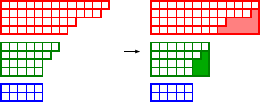}
\caption{Arrange rows into groups of warp size and then pad accordingly. In this figure, we use a warp size = 4 only for purposes of illustration.}
\label{fig:pad}
\end{figure}

\begin{figure}[htbp]
\centering
\includegraphics[width=.7\textwidth]{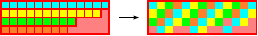}
\caption{Reorder the first warp in Figure \ref{fig:pad} in a coalesced column-major ordering.}
\label{fig:column-major}
\end{figure}

The proposed kernel algorithm is based on ELL. To mitigate the observation of row length distribution irregularity, we first sort the rows by length from longest to shortest (Figure \ref{fig:sort}).
A thread per row execution scheme is used, however, instead of using the representation used in the ELL format where padding continues until the maximum matrix row length is reached, each warp is now padded  up to the maximum row length of given warp (Figure \ref{fig:pad}).
To ensure coalesced memory access, we permute the data within each warp in a ``column-major" ordering similar to MGPU and ELL/HYB (Figure \ref{fig:column-major}).
Lastly, the results are mapped back to the original row-ordering when writing results to the solution vector $\vec{y}$. This algorithm will be referred to as ELL-WARP (K1). The code is shown in Listing \ref{lst:warpkernel1}. K1 requires that a list of values and their column indices be padded per warp, which are then sorted by row length and arranged in ``column-major" order. A list of the initial offsets for each warp is given as well as a row mapping from the sorted row index to the  original row index.

\begin{figure}[htbp]
\begin{lstlisting}[language=C++,
caption={Code for K1.},
label=lst:warpkernel1]
template <uint WARP_SIZE,bool usecache>
__global__ void K1( *A, *colinds, *Pinv, maxrows, *x, *warp_offset, *y, nrows) {

  const uint tid = threadIdx.x;
  const uint id = tid  + blockIdx.x * blockDim.x;
  const uint wid = tid & (WARP_SIZE-1);
  const uint warpid = id / WARP_SIZE;

  if (id < nrows) {

    IndexType toffset = warp_offset[warpid] + wid;
    uint maxnz = maxrows[warpid] * WARP_SIZE + toffset;

    // Perform sequential sum
    ValueType sum = A[toffset] * cache<usecache> (colinds[toffset],x);
    for(toffset += WARP_SIZE; toffset < maxnz; toffset += WARP_SIZE) {
      sum += A[toffset] * cache<usecache> (colinds[toffset],x);
    }

    // Store remapped result
    y[Pinv[id]] = sum;
  }
}
\end{lstlisting}
\caption{A is an array that contains the entries to the matrix. colinds are the index of the columns. Pinv is the inverse permutation array. x is the x vector. maxrows is an array which contains the maximum row length of a particular warp. warp\_offset determines the index at which the first thread of a warp should load from A and colinds. y is the solution vector. nrows is the total number of rows in the matrix. tid is the thread index. id is the global row id corresponding to a particular thread. wid is the index of a thread within a warp. warpid is the global index of a particular warp. toffset is the offset with respect to warp\_offset that each thread should load values from A and colinds. maxnz is the used to properly bound the workset of a particular thread. WARP\_SIZE and usecache are template variables. WARP\_SIZE sets the size of a warp for the given GPU architecture, while usecache determines whether the texture cache is used or not in loading from the x vector.}
\end{figure}

The purpose of this arrangement is to exploit the relatively small variances in row lengths per warp to reduce unnecessary padding in ELL and also to increase the amount of threads performing meaningful operations. As values and column indices are zero-padded, those values and column indices are cached and should not reduce the effectiveness of this algorithm if the variances within each warp are in fact small. By having a fixed row length for each warp, one can reduce the amount of data needed to compute row sums in comparison to CSR and COO while reducing the number of trivial instructions executed for padded entries in comparison to ELL. Thus, this kernel aims to achieve an equally-distributed workload similar to MGPU in cases where there is some degree of row length regularity within each warp. Unfortunately, the final non-coalesced memory write to the solution vector is a potential inefficiency for this kernel which is otherwise coalesced in terms of memory access to values and column indices.

\subsection{ELL-WARP with row reordering (K1r)}
While reordering the rows of the matrix helps us achieve row length regularity within individual warps, a straightforward SPMV operation with the row permuted matrix will yield a permuted solution.
\begin{equation}
A_{i,j}x_{j} = y_{i} \qquad A_{p(i),j}x_{j} = y_{p(i)}
\end{equation}
Thus the final memory store operation undoes this permutation, $p(i)$, such that the original ordering of the solution is preserved; however this operation results in a non-coalesced memory storage pattern. To mitigate the necessity of performing a non-coalesced write within our kernel, the column indices of the matrix and the vector $\vec{x}$ can be reordered such that the numbering of the corresponding entries in $\vec{x}$ and the solution vector are consistent. This is commonly performed in finite element computations by renumbering the unknowns during the setup phase of most codes. Reordering the columns of the matrix and entries of $\vec{x}$ does not change the result of the solution vector.
\begin{equation}
A_{i,j}x_{j} = y_{i} \qquad A_{i,f(j)}x_{f(j)} = y_{i}
\end{equation}
By performing this reordering, we can essentially ignore the original row mapping necessary in K1 as the sorted solution vector is now numbered consistently with the newly arranged $\vec{x}'$ vector. Unfortunately, the reordering of the column indices now produces a non-ordered access pattern when reading values from the $\vec{x}'$ vector. While the effects may be mitigated by using a texture cache, this is somewhat undesirable.

To further illustrate the reordering scheme, we considered a CSR ordered matrix for convenience. If $x^{\scas{num}}$ is the original ordering for $\vec{x}$, then $\ten{P}$ is the new ordering such that the rows will be sorted accordingly from longest to shortest.
\[
\begin{array}{r|c}
x^{\scas{num}} & \begin{bmatrix}
1 &2 &3 &4 &5 &6 &7
\end{bmatrix} \\
\hline
$\ten{P}$ & \begin{bmatrix}
2 & 5 & 7 & 3 & 1 & 4 & 6
\end{bmatrix} \\
\end{array}
\]
The reordered vector, $\vec{x}'$, and the reordered column indices, $\vec{c}'$, are defined below, where $\ten{P}$ is a permutation operator:
\begin{align}
\vec{x}'(\cdot) & = \vec{x}(\ten{P}(\cdot)) \\
\vec{c}'(\cdot) & = \vec{c}(\ten{P}^{-1}(\cdot)) \\
(\cdot) & = \ten{P}(\ten{P}^{-1}(\cdot))
\end{align}
Given the following single-row CSR matrix $\ten{A}$, a dense vector of column indices $\vec{c}$, and a dense vector $\vec{x}$
\begin{gather*}
\ten{A} = \begin{bmatrix}
7 & 8 & 9 & 10 & 2
\end{bmatrix}
\\
\vec{c} = \begin{bmatrix}
1 & 2 & 4 & 5 & 6
\end{bmatrix}
\\
\vec{x} = \begin{bmatrix}
1 & 2 & 3 & 4 & 5 & 6 & 7
\end{bmatrix}
\end{gather*}
the K1r scheme described above will re-arrange the data such that the following results.
\begin{gather*}
\vec{c}' = \begin{bmatrix}
5 & 1 & 6 & 2 & 7
\end{bmatrix}
\\
\vec{x}' = \begin{bmatrix}
2 & 5 & 7 & 3 & 1 & 4 & 6
\end{bmatrix}
\end{gather*}
This reordering scheme does not require that $\ten{A}$ be ordered, but results in non-ordered column indices, $\vec{c}'$, which is possibly detrimental to maintaining ordered access from $\vec{x}'$.

\subsection{K1r with sorted column index numbering (K1rs)}

The following is a variation on K1r that attempts to fix the issue with non-ordered access to the $\vec{x}'$ vector. The solution is to renumber the column indices of the matrix A. This requires a reordering of the values of the matrix in each row such that the reordering of the column indices is sorted in order after the reordering of the vector $\vec{x}$.

Again using the previous single-row CSR matrix as an example, we sort $\vec{c}'$ such that it is ordered from the smallest to largest indices within each row.
\[
\vec{c}{''} = \begin{bmatrix}
1 & 2 & 5 & 6 & 7
\end{bmatrix}
\]
\[
\ten{A}{''} = \begin{bmatrix}
8 & 10 & 7 & 9 & 2
\end{bmatrix}
\]
The vector $\vec{x}'$ remains the same; however, now it is necessary to sort the values of $\ten{A}$ within each row to properly account for the changes in $\vec{c}{'}$. The result is a intra-row reordered CSR matrix, $\ten{A}{''}$, and an ordered set of corresponding column-indices, $\vec{c}{''}$.

\subsection{ELL-WARP v2 (K2)}
To accommodate the possibility of larger intra-warp row length variances, a multiple threads per row variation of K1 is proposed. For example, if there exists a small number of abnormally long rows, it may be advantageous to process those rows with more threads while allowing the rest of the rows in the matrix to be processed by K1.

\begin{figure}[htbp]
\includegraphics[width=.9\textwidth]{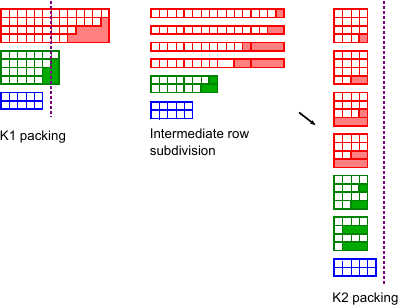}
\caption{A threshold value (purple dashed line) is given and the original padding in K1 is reorganized into smaller more evenly distributed warps.}
\label{fig:wpk2}
\end{figure}

A threshold value for the maximum number of entries a thread should be assigned is prescribed. If a row does not meet this criterion, the row is subdivided in half recursively until it meets this requirement. However, to avoid explicit thread synchronization within a block, if a row is longer than 32 $\times$ threshold, one warp will at the least process the entire row to maintain warp-to-warp row independence. The subdivision and repacking of the matrix is shown graphically in Figure \ref{fig:wpk2}.

\begin{figure}[htbp]
\begin{center}
\includegraphics[width=0.4\textwidth]{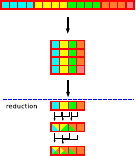}
\end{center}
\caption{Coloring represents the data for a particular thread. In this instance, 4 threads are used to process the representative row. Data is arranged in column-major order, and then a parallel reduction step occurs resulting with the final sum in the first entry of the output row.}
\label{fig:seg_redux}
\end{figure}

The variation inevitably increases the amount of padding in comparison to K1 based on the threshold value chosen. However, it allows unusually long rows to be processed more efficiently, while smaller rows retain the same padding as K1. The threshold parameter forces almost all warps to process the same number of non-zero entries. The kernel now includes a simple parallel reduction loop, which reduces rows in a given warp efficiently using shared memory. This is seen in Figure \ref{fig:seg_redux} for the first row of the representative warp in our illustrative example.

As rows are sorted initially, it is relatively easy to reprocess the original warp-sized row groupings from K1 and subdivide the numbers of rows per warp as needed to meet the threshold criterion. Extra information per warp is necessary to dictate the row number per warp and also the number of reductions needed per warp. However, this extra indexing information is minimal and results in two extra integers retrieved from global memory per warp. The code for ELL-WARP v2 (K2) is shown in Listing \ref{lst:warpkernel2}.

\begin{figure}[htbp]
\begin{lstlisting}[language=C++,
caption=Code for K2,
label=lst:warpkernel2]
template <uint WARP_SIZE,bool usecache>
__global__
void K2(ValueType* A, IndexType *colinds,
		    IndexType *rowmap, uint* maxrows,
		    ValueType* x, IndexType *warp_offset,
		    ValueType* y, IndexType nrows,
		    uint* reduction, uint* rows_offset_warp,
		    int nwarps) {

  const uint tid = threadIdx.x;
  const uint id = tid  + blockIdx.x * blockDim.x;
  const uint wid = tid & (WARP_SIZE-1);
  const uint warpid = id / WARP_SIZE;

  extern volatile __shared__ ValueType sumvalues[];

  if (warpid >= nwarps) return;
  const uint offsets = reduction[warpid];
  const uint row_start = rows_offset_warp[warpid];
  const uint rowid = row_start + wid/offsets;

  if (rowid < nrows) {

    IndexType toffset = warp_offset[warpid] + wid;
    const uint maxnz = maxrows[warpid] * WARP_SIZE + toffset;
    ValueType sum = A[toffset] * cache<usecache> (colinds[toffset],x);

    for(toffset += WARP_SIZE; toffset<maxnz; toffset += WARP_SIZE) {
      sum += A[toffset] * cache<usecache> (colinds[toffset],x);
    }
    sumvalues[tid] = sum;

    for (int i = 1; i< offsets; i <<= 1) {
      if (offsets > i ) {
	sum += sumvalues[tid+i];
	sumvalues[tid] = sum;
      }
    }

    if ((wid & (offsets-1)) == 0) {
      y[rowmap[rowid]] = sum;
    }
  }
}
\end{lstlisting}
\caption{Most of the pseudo-code terms are explained in Listing \ref{lst:warpkernel1}. reduction determines how many stages of parallel reduction should be used for a given warp. An offset of 1 means there are 0 stages of reduction, while an offset of 32 means that there are 5 stages of reduction and all threads in a warp are working on one row. rows\_offset\_warp determines the starting global row index of a warp. row\_id is the global row index a particular thread will contribute results to. nwarps is the number of warps needed to compute the SPMV.}
\end{figure}

\subsection{K2 variations (K2r) and (K2rs)}
Lastly, to avoid the final row mapping from sorted to non-sorted rows, the same variations proposed for K1 can be simply applied directly to K2. The fundamental difference between K1 and K2 is in how the matrix is repackaged into blocks of data. In K2, the packaging is more regular within blocks, whereas in K1, the packaging is more regular only within a given warp.

\section[Benchmarks of SPMV Methods]{Benchmarks of Sparse-matrix Vector Multiplication Methods}
Our novel algorithms above are compared against several available standard SPMV algorithms using a set of sparse matrices commonly used in benchmarking SPMV methods \cite{Williams2009178,matrix_bench,bell09}. Since the K1 and K2 kernels were developed for finite element simulations of the heart, we have also included 3 different refinements of a  patient-specific heart mesh as part of the set of benchmark matrices. General information about each of the benchmark matrices is included in Table \ref{tab:matinfo}.
%
\begin{table}[htbp]
\caption{Matrix benchmark information. The number of nonzeros (nz), the number of rows (nrows), bytes, minimum row length (minrow), maximum row length (maxrow), and average number of non-zero values per row (nz/nrows) is reported for each benchmark matrix. The matrices highlighted (\textbf{bold}) are finite element meshes of different refinements of a patient-specific heart mesh.}
\begin{center}
\input{matrixinfo}
\end{center}
\label{tab:matinfo}
\end{table}
Matrix structure properties vary substantially from matrix to matrix. It is apparent that certain matrices, such as Circuit and Webbase, have a very skewed non-zero row length distribution. On the other hand, some matrices such as Epidemiology and QCD have very regular row length distributions.

The following kernels were then chosen for comparison: CUSP-CSR and CUSP-HYB \cite{Cusp}, CUSPARSE \cite{CUSPARSE}, MGPU \cite{mgpu}, K1/r/rs, and K2/r/rs. The published MGPU results report the kernel time by performing a large number of products $\ten{A} \, \ten{x}$ within a loop, and reporting the average runtime per product. It was observed that the performance of the MGPU kernel varies depending on the number of iterations, with better performance at a higher number of iterations. However, in practice, a single product $\ten{A} \, \ten{x}$ is computed followed for some vector operations, after which further products $\ten{A} \, \ten{x}$ are needed. Therefore it is unclear to what extent the performance of the MGPU kernel reported for a large number of products $\ten{A} \, \ten{x}$, computed immediately after one another, is relevant. Therefore, we will show results for one run of the kernels here, while the results for 1200 iterations can be found in the appendix (Section \ref{sec:appendix}).

\begin{figure}[htbp]
\begin{center}
\includegraphics[width=\textwidth]{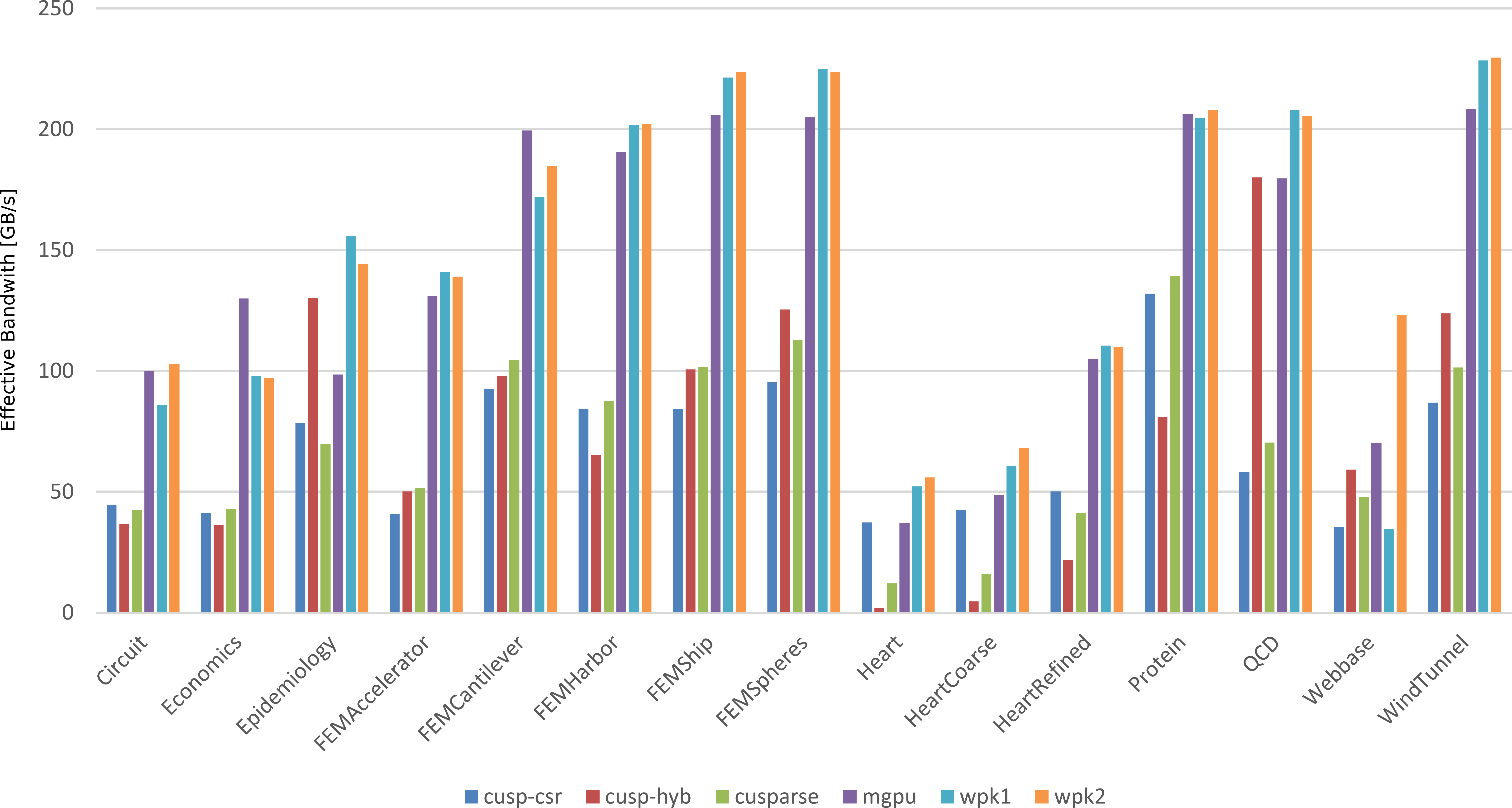}
\end{center}
\caption{Effective bandwidth benchmark results over matrices for 1 run are shown for CUSP-CSR (blue), CUSP-HYB (red), CUSPARSE (green), MGPU (purple), K1 (cyan), and K2 (orange).}
\label{fig:data1}
\end{figure}

For each matrix used in the benchmark, kernel times are averaged over a specified number of iteration runs. For the K1/r/rs kernels, different block sizes are varied from 32 to 256 threads by increments of 32. Likewise the same is done for the K2/r/rs kernels except, in addition, different maximum non-zero thresholds are varied from the minimum row length of the matrix to the maximum row length of the matrix. Lastly for MGPU, the following number of values per thread are used: 4, 6, 8, 10, 12, and 16. After the parameters for K1, K2, and MGPU are acquired, the fastest averaged times are found. The effective bandwith is reported in Figure \ref{fig:data1} and is defined as the rate of bytes processed from the matrix data by a kernel over time.

All benchmarks were run on a single Asus ENGTX480 graphics card (CUDA Compute compatibility 2.0) and on a PC with an Intel I7 950 CPU and 12 GB of memory. Kernels are compiled with optimizations enabled (--O3) and kernels use the texture cache when possible. The results for a single iteration run is shown in Figure \ref{fig:data1}. The parameters for MGPU, K1 and K2 are shown for each of the best performing kernels for each matrix in Table \ref{tab:best}.

The results for a single iteration indicate that our K1 and K2 algorithms perform well overall. For our patient-specific heart meshes, K1 and K2 algorithms perform better than the other tested algorithms. MGPU performs best for 2 matrices: Economics, and FEMCantilever. Meanwhile, K1 and K2 algorithms perform well even for Economics and FEMCantilever, and in general provide the best performance over the benchmarked matrices. Lastly, the effective bandwidth results vary slightly with the number of iterations for MGPU and CUSPARSE, where an increase in the number of iterations improves performance. These results can be found in the Appendix \ref{sec:appendix}. Despite these variances, the K1 and K2 kernels still outperform or are at least comparable to MGPU.


\begin{table}[htbp]
\caption{Benchmark best parameters for MGPU, K1 and K2.}
\begin{center}
\input{bestvalues}
\end{center}
\label{tab:best}
\end{table}

In many ways it is quite surprising that a relatively simple kernel algorithm (22 lines) has comparable performance to a sophisticated segmented scan algorithm in the case of MGPU (145 lines). Unlike other GPU SPMV algorithms, K1 and K2 are monolithic kernels, and extra kernel invocations are unnecessary. However, the thread block size for K1 and K2 are generally different, and thus their parameters must be found individually.

The effect of warp-level organization vs. global data restructuring in HYB can be inferred from Figure \ref{fig:data1} since K1 and K2 are partially warp-based variations of the ELL SPMV implementation and thus related to the HYB format. The results from the two variations of ELL show drastically different performance results for the majority of the benchmarked sparse matrices. K1 and K2 reduce the amount of padding with respect to ELL and simultaneously reduce the amount of memory transactions needed for computing the SPMV operation in comparison to HYB, thereby providing a dramatic increase in performance. K2 is used when there is a large difference in row length regularity and effectively handles outlier rows in a hybrid ELL-CSR like manner. Together, K1 and K2 are substantially better than ELL and HYB for sparse matrices.

\subsection*{Cost of Reordering Matrix Values}
To determine whether it is beneficial overall to reorder the data in a ``column-major" coalesced pattern in SPMV applications for finite elements, we consider the following. If K1 and K2 are used to calculate sparse matrix vector multiplications in the conjugate gradient method, only one initial transpose of values is necessary at the beginning of each Newton-Raphson iteration with the assumption that the assembler passes a CSR formatted matrix to the solver. Column indices do not need to be reordered, as we assume that the connectivity of the Lagrangian mesh does not change during the simulation; therefore, only the values of the tangent matrix change while the matrix structure remains constant. We can then determine the number of CG iterations necessary, such that K1 and K2 will outperform CUSPARSE by the following
\begin{equation}
t_{\scas{reorder}} + \alpha t_{\scas{wpk}}  \leq \alpha t_{\scas{cusparse}}
\end{equation}
where $\alpha$ is the number of iterations necessary such that a reordering of data from CSR to ``column-major" provides a benefit for the CG solver. $t_{\scas{reorder}}$ is the time to reorder the CSR tangent matrix into a coalesced K1- and K2-compatible form. $t_{\scas{wpk}}$ and $t_{\scas{cusparse}}$ are the SPMV operation times for K1/K2 and CUSPARSE respectively. The following data is used to compare the differences between reordering on the GPU and on the CPU. As our algorithm produces a reordering mapping for all non-zero entries during the initial scan, scatter operations are simply performed on the CPU using a simple for-loop, and the thrust::scatter() is used on the GPU \cite{Thrust}.
\begin{table}[htbp]
\caption{Comparison of $\alpha$ ratios for the GPU and CPU for K1 and K2. $\alpha$ represents the number of SPMV operations necessary for the benefit of K1/K2 kernels to be apparent over traditional non-reordered SPMV algorithms. The different refinements of our patient specific meshes are \textbf{bolded}. $\alpha = \infty$ means that for that particular matrix the cost of reordering a CSR matrix into a K1/K2-compatible form cannot be shown because $t_{\scas{wpk}}$ is slower than $t_{\scas{cusparse}}$.  }
\begin{center}
\input{tabalpha}
\end{center}
\label{tab:alpha}
\end{table}
From Table \ref{tab:alpha}, it is fairly clear that CPU reordering is substantially slower than reordering directly on the GPU even for small matrices (Heart, HeartCoarse, etc...). On average, GPU reordering is 5 times faster than CPU reordering over all of the tested matrices. GPU reordering still takes longer than the GPU SPMV kernel for the majority of matrices. However, even when reordering on the CPU at every Newton-Raphson iteration for the finite element benchmark matrices, $\alpha_{\scas{CPU}}$ is within the general number of iterations for most CG problems. On the other hand, GPU reordering is very fast, and the benefits should be noticeable within 10 iterations for both K1 and K2 on average. Thus, for the remaining number of iterations, the benefit of K1 and K2 over other kernels will be evident.

\begin{figure}[htbp]
\begin{center}
\includegraphics[width=\textwidth]{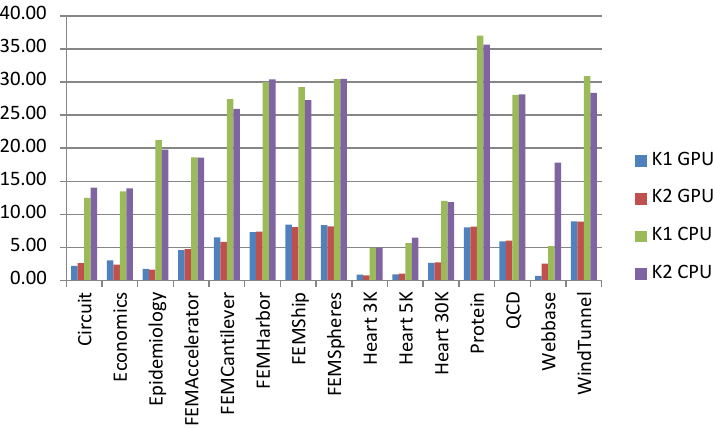}
\end{center}
\caption{Factor of reordering time to kernel time ($t_{\scas{reorder}}:t_{\scas{wpk}}$) on the GPU and CPU for K1 and K2 kernels are shown. Factors for K1 for the GPU are shown in blue, and for the CPU are shown in green. For K2, GPU reordering is shown in red, and in purple for the CPU.}
\label{fig:reorderratio}
\end{figure}

From Figure \ref{fig:reorderratio}, one initially may discount the benefit of using any reordered kernels for global finite element tangent assembly. However, the ordering for finite element assembly can be arranged, such that the resulting ordering is already ``column-major" ordered. Likewise the result of the assembly of the tangent matrix can also be arranged such that results are already coalesced and ordered properly; thus, bypassing the need for reordering the matrix $\ten{A}$ altogether. Only, in this way, can the performance of the synthetic benchmark results for K1 and K2 be obtained for finite element assembly in real applications.

\section{Effect of individual optimizations}
In this section, we evaluate different aspects of performance. We first investigate whether different kernel level optimizations and the variations to K1 and K2 provide an improvement in performance. The cost-effectiveness of GPU data reordering is then compared and evaluated against CPU data reordering implementations.

A subset of the benchmark matrices were chosen to determine the effects of different kernel optimizations taken in the development of K1 and K2 and their variants. The matrices chosen were Circuit, Epidemiology, FEMHarbor, Heart 3K, Heart 5K, Heart 30K, and QCD. The following factors were measured for an ELLWARP algorithm (KX): coalesced vs. non-coalesced memory access patterns, sorted vs. unsorted rows, kernels that involve remapping vs. those that renumber column indices and reorder the vector $\vec{x}$ (KX vs. KXr), and lastly the differences between two possible column numbering schemes (KXr vs. KXrs). See Table~\ref{tab:optimizations} for a summary of the acronyms used.

\begin{table}[htbp]
\begin{center}
\begin{tabular}{|c|p{0.6\textwidth}|}\hline
Kernel Approach	& Description \\ \hline
KX	& Matrix element data is divided into sorted warp-sized segments and reordered with ``column-major'' ordering. \\ \hline
KXr	& In addition to KX, $\vec{x}$ is reordered and the column indices of the matrix are renumbered accordingly to avoid non-coalesced assignment.\\ \hline
KXrs & In addition to KXr, the column indices are renumbered such that access to the $\vec{x}$ vector is ordered. The matrix values are subsequently rearranged in a corresponding manner.\\
 \hline
\end{tabular}
\end{center}
\caption{Description of the three different approaches to sorting and renumbering a kernel X, which is either kernel K1 or K2.}
\label{tab:optimizations}
\end{table}

\begin{figure}[htbp]
\begin{center}
\includegraphics[width=0.8\textwidth]{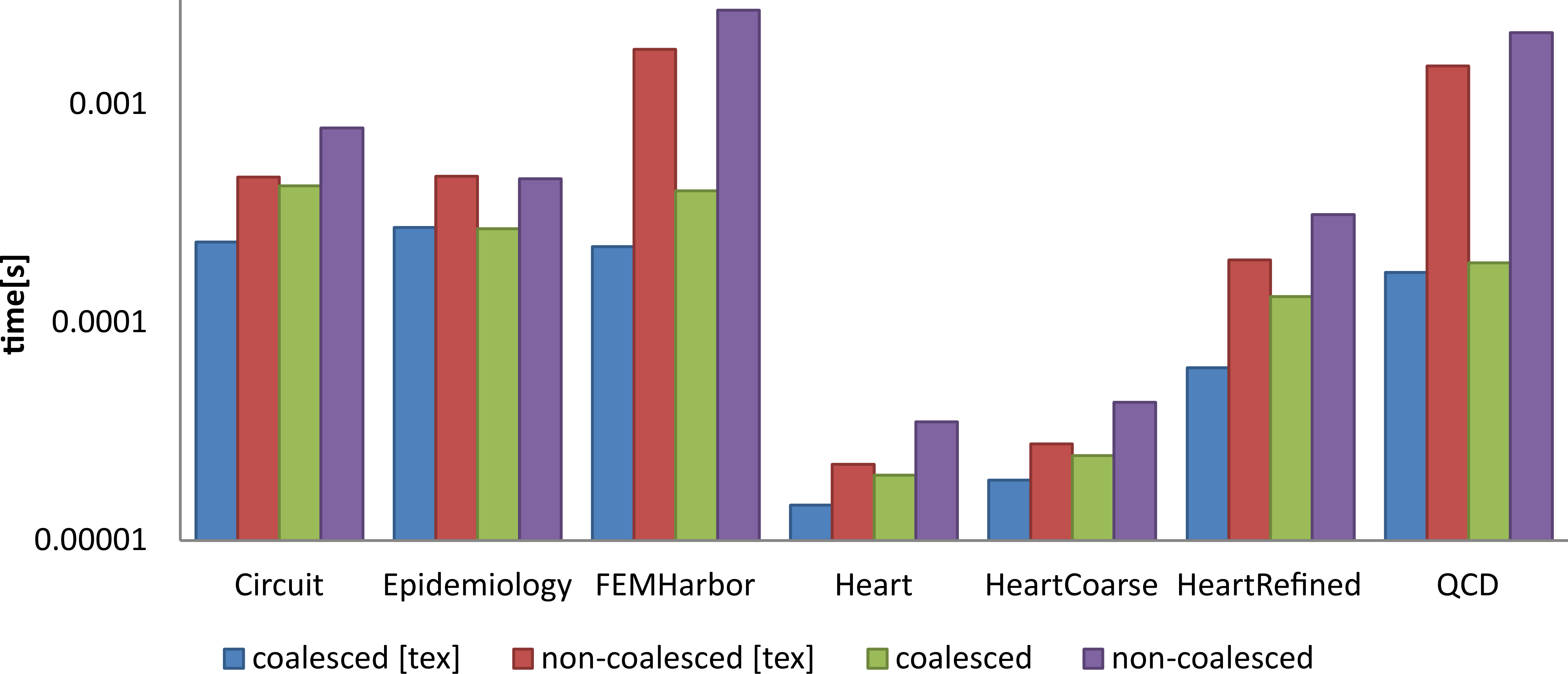}
\end{center}
\caption{Comparison of coalesced ``column-major" data vs. non-coalesced data ordering kernel times for K1.}
\label{fig:compcoal}
\end{figure}

First, we investigate the importance of coalesced memory access in Figure \ref{fig:compcoal}. In this test, we compared our K1 kernel against a kernel where data was left in the standard CSR ``row-major'' form instead of a ``column-major'' ordering that results in coalesced memory access within each warp.
As expected, coalesced memory access patterns result in a dramatic reduction in computation time in comparison to non-coalesced memory access patterns. In fact, a coalesced memory access pattern accounts for a 1.5 to 9 fold increase in speed when using texture memory access, and for a 1.5 to 10 fold increase in speed without textures.

We also tested the effect of sorting matrix rows from longest to shortest (K1) and compared it to the equivalent algorithm where the ordering of rows of the original matrix and $\vec{x}$ vector are preserved (non-sorted). As the non-sorted implementations must result in extra zero-padding compared to the sorted ELLWARP implementations, the table below (Table \ref{tab:nosort}) reports the difference in padding and also shows the percentage of time-difference with respect to a non-row ordered algorithm for K1. The time-difference is defined as the K1 kernel time subtracted from the non-sorted K1 kernel time. The time-difference percentage is simply the percentage of time-difference with respect to the non-sorted kernel time.

\begin{table}[htbp]
\caption{Benchmark data regarding the effects of sorting for K1 are shown below. The padding difference percentage designates the percentage of padding that differs between the padded sorted and non-sorted rows with the padded non-sorted representation as reference. A positive padding difference percentage means that the sorted K1 kernel reduces padding. The time difference percentage is defined as the time difference, the K1 kernel time subtracted from the non-sorted kernel time, divided by the non-sorted kernel time. A positive time difference means that the K1 kernel is faster when sorted.}
\resizebox{\textwidth}{!} {
\begin{tabular}{|l|c|c|c|c|c|c|c|}
\hline
& Circuit	&Epidemiology	&FEMHarbor&	Heart	&HeartCoarse&	HeartRefined&	QCD \\
\hline
Padding Difference Percentage & 59.89\% &	0.14\%	&33.67\%&	29.60\%&	33.73\%&	30.03\%&	0\\
Time difference (textured) & 21.16\% &	-2.10\% &	0.00\% &	3.00\% &	22.67\% &	5.79\% &	0.47\% \\
Time difference & -17.38\%	&-4.35\%	&0.00\%	&-28.22\%	&10.42\%	&-46.80\%	&0.49\% \\ \hline
\end{tabular}
}
\label{tab:nosort}
\end{table}
The effect of sorting on performance is not entirely clear. For the majority of matrices, sorting the matrices resulted in an increase in performance when using the texture memory cache; however sorting may have detrimental effects on other matrices as shown for Epidemiology even when using the texture cache. On the other hand, without texture cache access, preserving the original row ordering results in significantly faster kernel times. Except for the Epidemiology matrix, the texture cache helps significantly, and the row-sorted texture cache kernel times are faster than the non-sorted kernel times without texture cache. However, from Table \ref{tab:nosort}, it is evident that Epidemiology and QCD have very similar row length regularity, and therefore the effects of sorting may actually disrupt the ordered access pattern for $\vec{x}$.  Texture access is therefore not as helpful in those cases. Overall, sorting the rows greatly reduces the zero-padding necessary, which is important in terms of applying SPMV operations in real situations.

Other tests were performed to compare the possible benefits from using the different kernel variations K$(\cdot)$r and K$(\cdot)$rs and their results can be found in the Appendix \ref{sec:appendix}. From the results, the performance improvement gained from reordering $\vec{x}$ to match the sorting of rows by length can be directly attributed to the avoiding the non-coalesced assignment required in the original kernel. However, each of the variations only provides a small speed-up in comparison to the preceding effects of coalesced ordering for the matrix and longest-to-shortest reordering.

Overall, our results show that column-major coalesced memory access, use of textures, and sorting are very important. This corroborates key insights made in developing the ELL, HYB, and MGPU SPMV algorithms \cite{bell09, mgpu} where column-major ordering can help achieve optimal memory throughput while retaining a degree of flexibility with regards to the number of threads and matrix values per row. The effect of using the texture cache in accessing values from $\vec{x}$ is significant in cases where access to the $\vec{x}$ is not already highly ordered, as in the case of Epidemiology. Lastly, while the effects of sorting rows from longest-to-shortest with the texture cache can be significant, the main benefit of the sorting is to reduce the amount of zero-padding which ultimately allows the K1 and K2 kernels to perform well even on larger highly unstructured meshes.

While the different variations provided additional benefits over the original K1 and K2 algorithms, the effects are less significant in comparison to the effects of column-major ordering and the use of the texture cache. Unfortunately, embedding alternate renumbering schemes adds some additional complexity to the finite element implementation. Therefore, if every bit of performance must be obtained, one would ideally use a pre-reordered $\vec{x}$ variation (Kr, Krs). While the speedup gains are compounded between K vs. Kr and Kr vs. Krs, proper modifications should yield substantial improvements in SPMV computation speed. However, as general drop-in SPMV replacements, K1 and K2 seem to perform quite well for many applications.


\section{Finite Element Comparison Between GPU and Multi-core}

In the final set of results, we compare the performance of our finite element GPU framework against a multi-core PETSc implementation of the same code. For the GPU SPMV implementation, we use the best parameters found for K1 and K2 from Tables \ref{tab:wpk1best} and \ref{tab:wpk2best} for 50 iterations. For MGPU, we simply have chosen \texttt{valuesPerThread}=6 and 10 for reference (see Table~\ref{tab:mgpubest} for details). We look at 4 different successive refinements of the Heart mesh: 3k, 5k, 30k, and 50k nodes. The benchmark  results for the 50k mesh use the same parameters for K1 and K2 SPMV kernels as those found for Heart 30K. For simplicity, the GPU finite element assembly implementation uses CUSPARSE for SPMV to assemble the global tangent matrix and global residual vector. For the multi-core PETSc implementation, we used a Jacobi preconditioned CG solver when possible; however there were convergence issues with the simple Jacobi preconditioner for the 50k mesh. However, for the purposes of comparison, we also show the results of our PETSc implementation using the block Jacobi preconditioner for all 4 mesh refinements. In these results, we use a CPU based reordering for MGPU, K1, and K2. On the CPU, the following results are reported for the PETSc Jacobi and block preconditioners running on 1, 2, 3, and 4 cores. CPU-reordering of the matrix was used for these results.

%
\begin{table} [htbp]
\caption{Benchmark best MGPU valuesPerThread parameter. The number of repeated SPMV iterations is denoted in brackets.}
\begin{center}
\input{mgpubestvalues}
\end{center}
\label{tab:mgpubest}
\end{table}


%
\begin{figure}[htbp]
\begin{center}
\includegraphics[height=0.25\textheight]{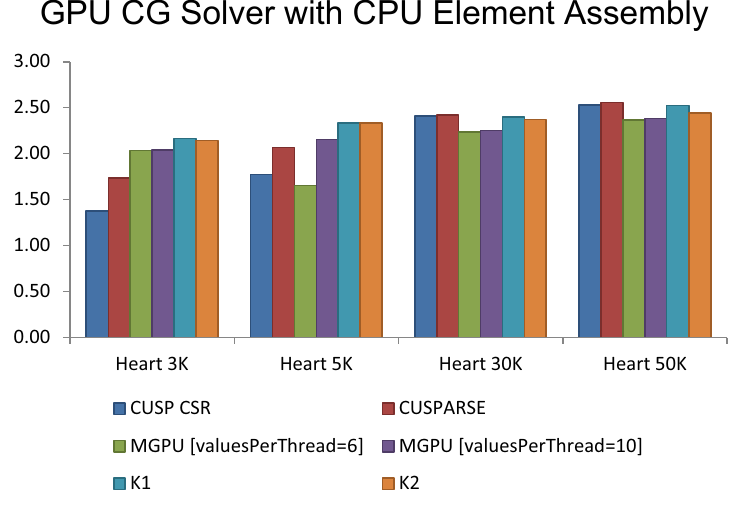}\\
\includegraphics[width=0.9\textwidth]{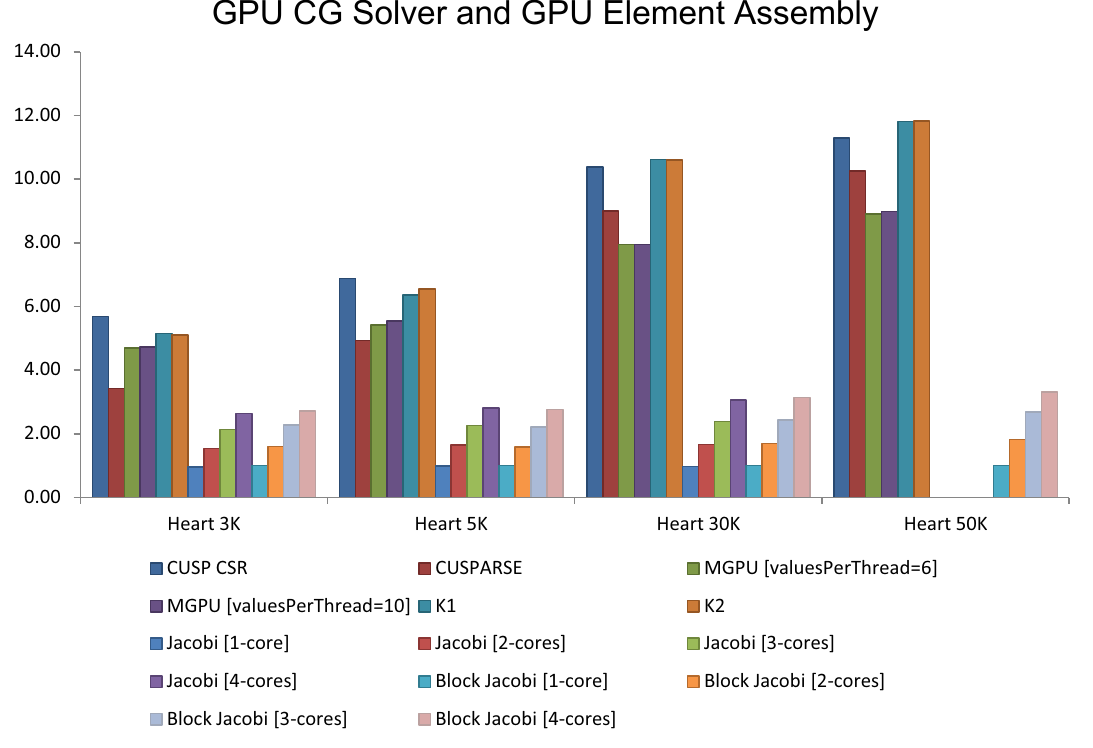}
\end{center}
\caption{The resulting factor of the increase in speed is shown for the GPU CG Solver and GPU finite element method. The factor of the increase in speed is defined as the ratio of the computation time of the single core PETSc block-jacobi CG solver compared to the computation time of the highlighted GPU SPMV algorithms denoted by the different colors. Results using a GPU CG solver with a standard single core CPU-based Element Assembly routine are shown (top). Results for GPU CG Solver and GPU Element Assembly implementations are shown (bottom) together with multi-core CPU PETSc implementation speedup results.}
\label{fig:speedups}
\end{figure}
The resulting speedup for the different tests with a single process PETSc block-jacobi solver with standard finite element assembly as reference is shown in Figure \ref{fig:speedups}. Several general observations can be made. First, the multi-core CPU PETSc implementation does not scale linearly. The block Jacobi preconditioner CG solver is slightly faster than the Jacobi preconditioner counterpart, but the parallel speed-ups on multiple cores are roughly the same. The GPU CG solver with CPU finite element assembly implementation is marginally slower than the 3-core PETSc implementation. The GPU finite element implementations are at least twice as fast compared to the quad process PETSc implementations.

Overall, the GPU CG solver implementations with finite element assembly start  with a two-fold speedup for the 3129-node mesh and increase to a factor of 2.5 for the 50,000 node heart mesh. CUSP-CSR and CUSPARSE-CSR seem to perform better in comparison to K1 and K2 as the number of nodes increases. It should be noted that with GPU-reordering the performance of K1 and K2 kernels improves slightly, however the CUSP-CSR and CUSPARSE-based CG solvers still initially perform better that the K1 and K2 kernels. This point will be clarified later in this section. MGPU also performs well, but is the poorest performing kernel in all cases. On the other hand, the fully-GPU finite element implementation for K1 and K2 ranges from a speedup factor of 5 to 12 as the mesh is subsequently refined. Again, CUSPARSE-CSR and CUSP perform better relative to MGPU as the mesh size increases; however, K1 and K2 kernels are only initially slower than CUSP-CSR. At larger mesh sizes of  30k and 50k nodes, K1 and K2 outperform CUSP-CSR even when using CPU data reordering.

The performance results found in the two finite element implementations do not seem to match those predicted by the synthetic benchmark. Namely, CUSP and CUSPARSE kernels scale significantly better than expected and in general outperform MGPU. It was found that K1 and K2 kernels actually outperform the best CUSP and CUSPARSE kernels results after adjusting the parameters to the K1 and K2 kernels empirically from the acquired best results for 50 iterations (Tables \ref{tab:wpk1best} and \ref{tab:wpk2best}) even without GPU reordering. This finding indicates that more work must be done in developing a better representative benchmark to determine the best parameters for K1 and K2 for finite element applications since the resulting best parameters for our finite element applications apparently do not match closely those found in our synthetic benchmarks.

\begin{figure}[htbp]
\begin{center}
\includegraphics[height=0.25\textheight]{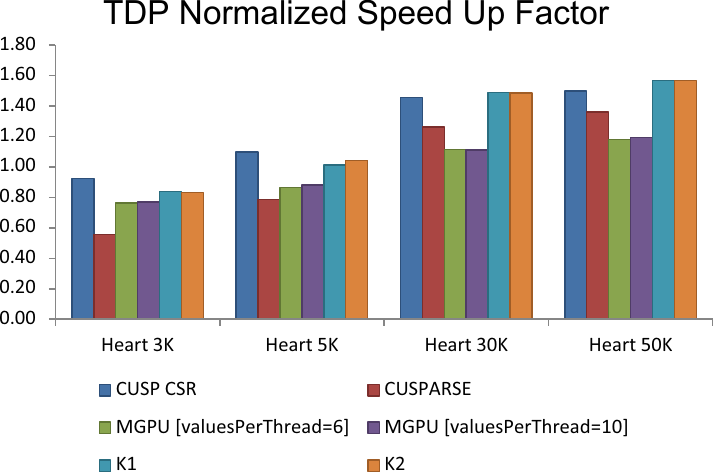}
\end{center}
\caption{The thermal design power (TDP) normalized resulting factor of increase in speed is shown for the GPU CG Solver and GPU finite element method with a 4-core PETSc block-jacobi CG solver as reference. }
\label{fig:normspeedups}
\end{figure}

We considered the power consumption as a measure of performance. The ASUS ENGTX480 GPU has a thermal design power (TDP) of 295W, while the I7 950 has a TDP of 130W. We normalized the speedup factor results to account for the difference in power consumption in Figure \ref{fig:normspeedups} to determine the computing effectiveness of our GPU algorithms as compared to computations on the CPU.

In the GPU CG/CPU assembly case (not shown in Figure \ref{fig:normspeedups}), we assume that the total CPU assembly and GPU CG solver implementation consume the maximum TDP of our CPU or GPU at worst. After power normalization, it was unfortunately found that the GPU CG/CPU assembly case is not power effective for any SPMV implementation at any mesh refinement chosen when compared against the 4-core parallel finite element CPU implementation. The K1 and K2 kernels performed the best and were able to achieve up to 60\% power effectiveness for the 4 mesh refinements.

However, for the fully-GPU implementations shown in Figure \ref{fig:normspeedups}, all implementations are power effective starting from the Heart 5K refinement and onwards. The power effectiveness starts at about 1.0 and increases to 1.6 for the K1 and K2 kernels. The CUSP CSR implementation performs comparably, but the differences between the K1 and K2 kernels and CUSP seem to widen as the problem size grows.

\section{Conclusion}
While the results have been very promising for our kernels, we believe these algorithms can be further improved both in terms of analysis and implementation of the algorithms in real applications. Reordering of values is still expensive compared to the actual SPMV computation time. Luckily it is possible to hide the reordering bottleneck within the assembly routine. This can be done by first reordering the location where each element stores its elemental stiffness and residual entries such that the kernel can directly use the stored matrix to generate a resulting global stiffness and residual matrix that is already reordered for the CG solver.

Since this particular physical problem is assembly heavy, improvements and optimizations to the finite element assembly operation will result in very substantial speed improvements in addition to improvements to SPMV operations. The shared-memory non-zero assembly operation \cite{cecka11} can be used to reduce global memory use and increase computational density. Coloring techniques may also be an alternative strategy in gaining additional performance in assembling the matrices from element matrices to global matrices. Another alternative is to leverage streaming algorithms to perform assembly operations and compute element quantities simultaneously to reduce global memory usage and potential race conditions.

Lastly, the slight differences between the optimal parameters for GPU CG solvers and the optimal parameters for GPU finite element implementations highlight potential avenues of improvement for our novel algorithms. Since the structure and access pattern of each mesh and matrix can be analyzed beforehand, it would be extremely beneficial to develop a metric for determining good partitioning parameters for the K1 and K2 algorithms a priori. Given the relatively simplicity of this algorithm, such a study should be possible, and would further increase the utility of these ELL-WARP algorithms.

In conclusion, we have shown how key insights from the ELL, HYB, pJDS and ModernGPU SPMV algorithms have led to the development of new, efficient SPMV algorithms that perform well over a large range of sparse matrices. The effects of different optimizations have been examined and ultimately lead to faster SPMV computation times. Lastly, this study highlights the potential use of GPUs for general finite element simulations.

\section {Acknowledgments} 
Part of this research was done at Stanford University, and was supported in part by the U.S.\ Army Research Laboratory, through the Army High Performance Computing Research Center, Cooperative Agreement W911NF-07-0027. This material is also based upon work supported by the Department of Energy National Nuclear Security Administration under Award Number DE-NA0002373-1. This work was also partly supported by the National Institutes of Health, grant U01 HL119578. J.\ Wong was also supported by the Sang Samuel Wang Stanford Graduate Fellowship.

\bibliography{paper}

\section{Appendix}
\label{sec:appendix}

The following sections have been included in the appendix to highlight interesting observations and results that support and add value to this paper, yet are not key findings to this work. The appendix is organized into three subsections: one section regarding the optimal parameters found for the MGPU, K1, and K2 kernels using our particular hardware, a second section comparing the differences between the kernel variations, and a final section where GPU reordering is utilized to improve the performance of K1 and K2 kernels in the context of SPMV operations.

\subsection{Optimal parameters for MGPU, K1, and K2 kernels}

\begin{figure}[htbp]
\begin{center}
\includegraphics[width=\textwidth]{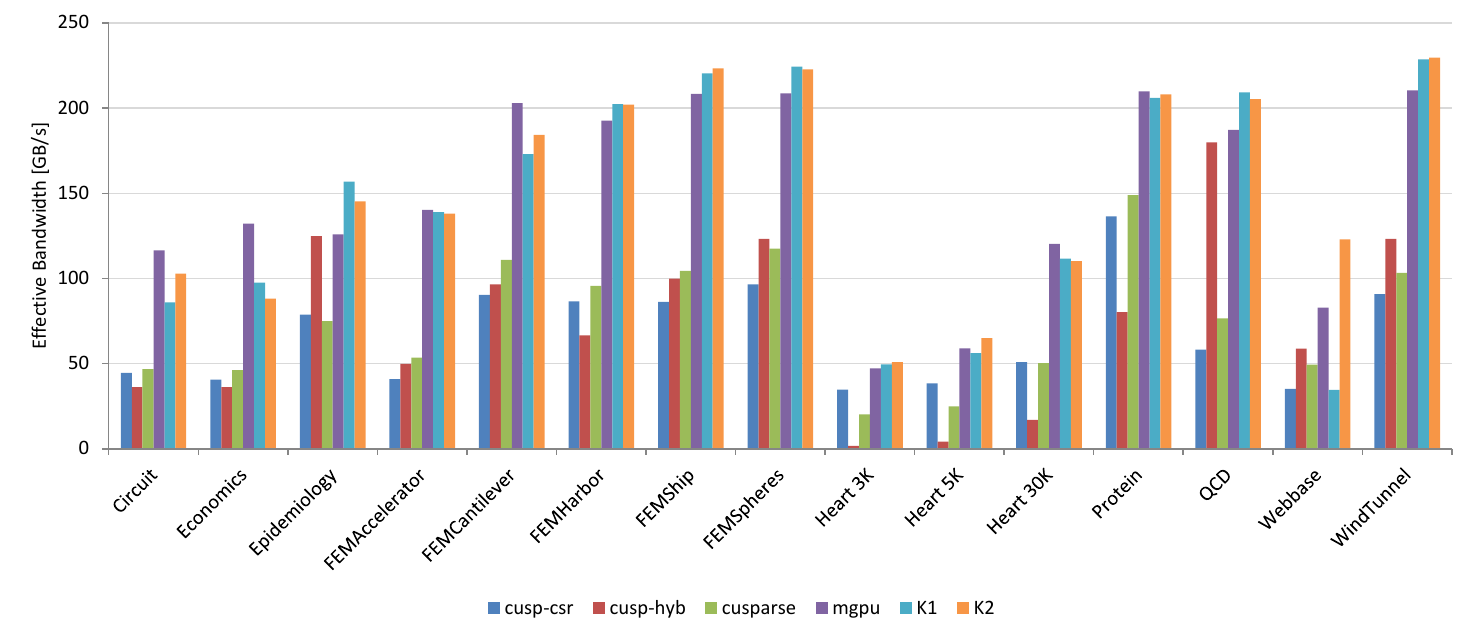}
\end{center}
\caption{Effective bandwidth benchmark results of matrices for 1200 iterations.}
\label{fig:data1200}
\end{figure}

Overall the results are fairly consistent. MGPU, K1, and K2 perform well over the benchmark sparse matrices. However, there are some slight differences between 1, 50, and 1200 iterations. The results for 1200 iterations are shown in Figure \ref{fig:data1200}. The performance of MGPU increases as the number of iterations increases. MGPU outperforms K1 and K2 kernels for 3 matrices at 1 iteration, but outperforms K1 and K2 kernels for 6 matrices at 1200 iterations. For the 3 matrices where MGPU gradually outperforms K1 and K2 kernels, the differences in performance are minimal. Overall, the performance of the K1 and K2 kernels are comparable to the performance of MGPU.

In general, K1 and K2 perform the best overall for all 15 matrices for 1, 50, and 1200 iterations. For a single iteration, the K1, K2 kernels are fastest for 13 matrices, while at 50 iterations, the K1 and K2 kernels are only fastest for 10 matrices. At 1200 iterations, a MGPU kernel is fastest for a total of 6 matrices: Circuit, Economics, FEMAccelerator, FEMCantilever, HeartRefined, and Protein. On the other hand, K1 and K2 kernels are fastest for a total of 9 matrices: Epidemiology, FEMHarbor, FEMShip, FEMSpheres, Heart, HeartCoarse, QCD, Webbase, and WindTunnel.


From Table \ref{tab:wpk1best}, we found that the K1 parameters rarely change between 1, 50, and 1200 iterations. K2 parameters are fairly constant, although a few matrices have threshold and block parameters that vary between 1, 50, and 1200 iterations (Table \ref{tab:wpk2best}). However, overall the results are fairly consistent across the range of iterations. The best performing MGPU kernels vary for several matrices, but are constant for others.

%
\begin{table} [htbp]
\caption{Benchmark best K1 blocksize parameter. The number of repeated SPMV iterations is denoted in brackets.}
\begin{center}
\input{wpk1bestvalues}
\end{center}
\label{tab:wpk1best}
\end{table}
%
\begin{table} [htbp]
\caption{Benchmark K2 best threshold and blocksize parameters. The number of repeated SPMV iterations is denoted in brackets.}
\begin{center}
\input{wpk2bestvalues}
\end{center}
\label{tab:wpk2best}
\end{table}


%

\subsection{Comparisons between kernel variations}
\begin{figure}[htbp]
\begin{center}
\includegraphics[width=0.8\textwidth]{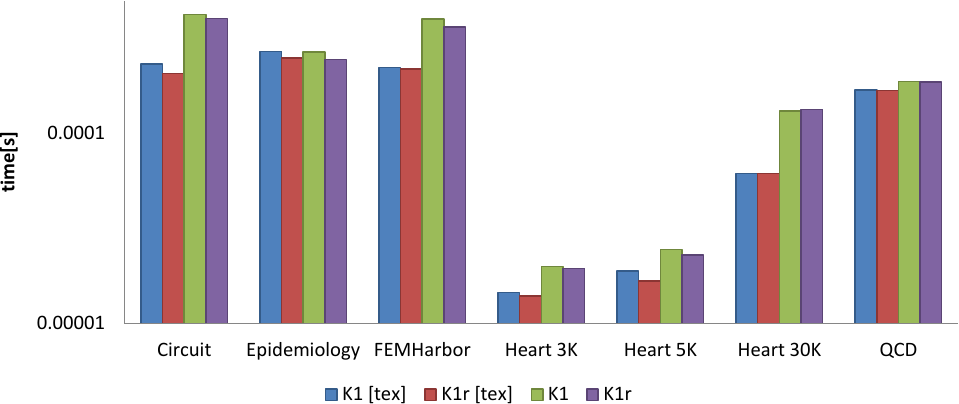}
\end{center}
\caption{Comparison of K1 vs K1r kernel times.}
\label{fig:compwpk1r}
\end{figure}

\begin{figure}[htbp]
\begin{center}
\includegraphics[width=0.8\textwidth]{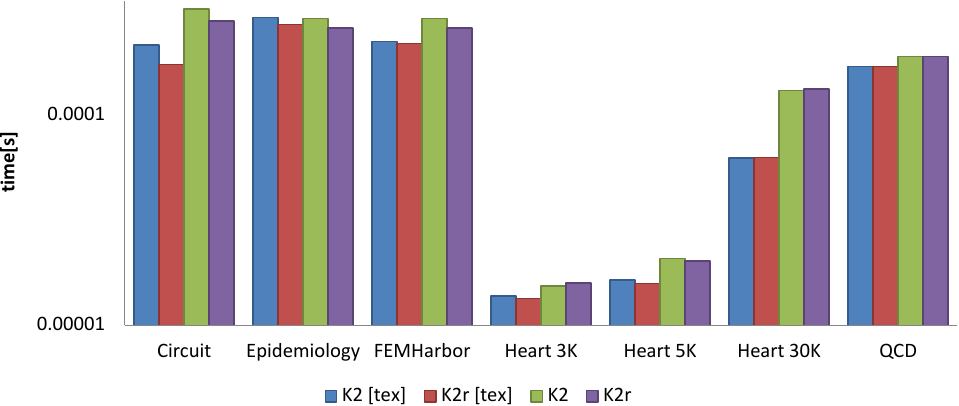}
\end{center}
\caption{Comparison of K2 vs K2r kernel times.}
\label{fig:compwpk2r}
\end{figure}

Next we examine the benefits of avoiding the final solution reordering necessary in K1 and compare it to a pre-reordered $\vec{x}$ and pre-renumbered $\vec{c}$. The results are shown in Figure \ref{fig:compwpk1r} and \ref{fig:compwpk2r}. The first reordered $\vec{x}$ kernel, K1r, outperforms K1 with texture cache access. Without the use of the texture cache, K1r outperforms K1, except in the case of the HeartRefined mesh. Likewise, with the texture cache enabled, K2r outperforms K2, but is sometimes slower without texture cache access. The effect of reordering $\vec{x}$ is generally beneficial, but is similar to the effect of sorting by row length where texture cache access for the well ordered matrices such as Epidemiology and QCD result in only slight differences between texture and non-texture cache results.

\begin{figure}[htbp]
\begin{center}
\includegraphics[width=0.8\textwidth]{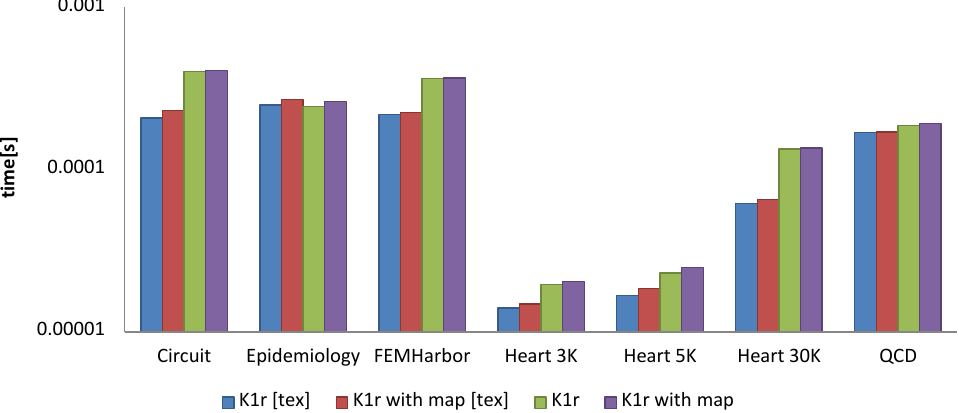}
\end{center}
\caption{Comparison of kernel times to investigate the cost of the final remapping to the original reordering vs K1r.}
\label{fig:compremap}
\end{figure}

To determine the cost of the final solution remapping, the K1r kernel was modified to perform the final solution remapping that results in a solution $\vec{y}$ that matches the original numbering for $\vec{x}$ and the results are reported in Figure \ref{fig:compremap}. In all cases, the cost of remapping is more than K1r. For K1r with use of the texture cache, the final solution remapping costs between 1\% to 11.5\% of the computation time of K1r. Whereas without the texture cache, remapping costs between 0.8\% and 8\% of the computation time of K1r. Similar results were found for K2r.

\begin{figure}[htbp]
\begin{center}
\includegraphics[width=0.8\textwidth]{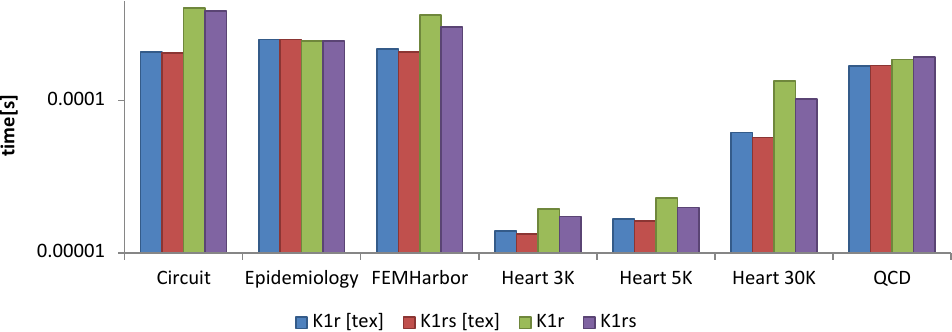}
\end{center}
\caption{Comparison of K1r vs K1rs kernel times.}
\label{fig:compwpk1rs}
\end{figure}

\begin{figure}[htbp]
\begin{center}
\includegraphics[width=0.8\textwidth]{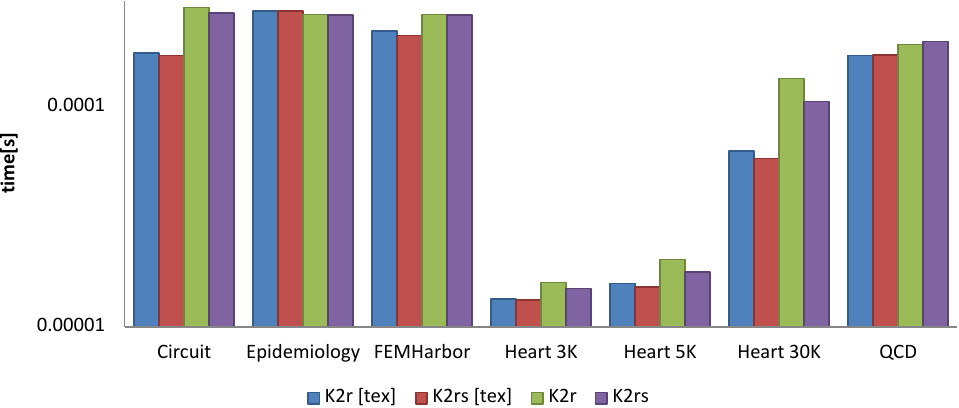}
\end{center}
\caption{Comparison of K2r vs K2rs kernel times.}
\label{fig:compwpk2rs}
\end{figure}

Next, we examine the improvement of K1rs over K1r and K2rs over K2r in Figures \ref{fig:compwpk1rs} and \ref{fig:compwpk2rs}. Other than the QCD and Epidemiology matrix cases which are already very well ordered, the values and row sorted versions of both kernels are marginally faster. K1rs is faster than K1r by 1.75\% to 7.25\% with the texture cache, and by 4\% to 23.5\% without the texture cache for matrices other than QCD and Epidemiology. K2rs is faster than K1r by 1\% to 7.5\% with the texture cache and 0.25\% to 21\% without the texture cache for matrices other than QCD and Epidemiology. K1rs and K2rs are slower than their Kr counterparts for the Epidemiology case by less than 0.25\% with and without the texture cache. For the QCD case, the row-sorted reordered kernels are slower by 0.25\% and 3.25\%.


\label{lastpage}
\end{document}

%% file: definitions.tex
\newcommand{\beq} {\begin{equation}}
\newcommand{\eeq} {\end{equation}}

\newcommand{\divg}{\mbox{\rm{div}}\,}

\def\sca   #1{\mbox{\rm #1}{}}

\def\vec   #1{\mbox{\boldmath $#1$}{}}
\def\ten   #1{\mbox{\boldmath $#1$}{}}
\def\scas  #1{\mbox{{\scriptsize{${\rm{#1}}$}}}{}}

%% file: matrixinfo.tex
\begin{tabular}{|c|r|r|r|r|r|r|}
\hline
MatrixName	&nz	&nrows	&bytes	&minrow	&maxrow	&nz/nrows\\
\hline
Circuit	&958,936	&170,998	&19,178,720	&1	&353	&6\\
Economics	&1,273,389	&206,500	&25,467,780	&1	&44	&7\\
Epidemiology	&2,100,225	&525,825	&42,004,500	&2	&4	&4\\
FEMAccelerator	&2,624,331	&121,192	&52,486,620	&8	&81	&22\\
FEMCantilever	&4,007,383	&62,451	&80,147,660	&1	&78	&65\\
FEMHarbor	&2,374,001	&46,835	&47,480,020	&4	&145	&51\\
FEMShip	&7,813,404	&140,874	&156,268,080	&24	&102	&56\\
FEMSpheres	&6,010,480	&83,334	&120,209,600	&1	&81	&73\\
\textbf{Heart} 3K	&\textbf{37,035}	&\textbf{3,129}	&\textbf{740,700}	&\textbf{5}	&\textbf{21}	&\textbf{12}\\
\textbf{Heart} 5K	&\textbf{52,715}	&\textbf{4,563}	&\textbf{1,054,300}	&\textbf{6}	&\textbf{22}	&\textbf{12}\\
\textbf{Heart} 30K	&\textbf{367,443}	&\textbf{28,639}	&\textbf{7,348,860}	&\textbf{6}	&\textbf{24}	&\textbf{13}\\
Protein	&4,344,765	&36,417	&86,895,300	&18	&204	&120\\
QCD	&1,916,928	&49,152	&38,338,560	&39	&39	&39\\
Webbase	&3,105,536	&1,000,005	&62,110,720	&1	&4700	&4\\
WindTunnel	&11,634,424	&217,918	&232,688,480	&2	&180	&54\\
\hline
\end{tabular}

%% file: bestvalues.tex
\begin{tabular}{|c|r|r|r|r|r|r|}\hline
MatrixName	&\multicolumn{1}{c|}{MGPU}	& \multicolumn{1}{c|}{K1}	& \multicolumn{2}{c|}{K2}\\
& valuesPerThread & blocksize & threshold & blocksize \\ \hline
Circuit	&6	&64 &80 &160 \\
Economics	&6	&64 &7 &96 \\
Epidemiology	&6	&128 &4	&128 \\
FEMAccelerator	&6	&64 &58	&64 \\
FEMCantilever	&10	&64 &19	&224 \\
FEMHarbor	&10	&96 &128	&96 \\
FEMShip	&10	&96 &28	&128 \\
FEMSpheres	&10	&96 &47	&128 \\
Heart 3K	&12	&32 &7	&256 \\
Heart 5K	&10	&64 &10	&192 \\
Heart 30K	&10	&64 &21	&64 \\
Protein	&10	&64 &172	&96\\
QCD	&10	&224 &39	&96\\
Webbase	&6	&32 &658	&128 \\
WindTunnel	&10	&96 &38	&96 \\ \hline
\end{tabular}

%% file: tabalpha.tex
\begin{tabular}{|c|r|r|r|r|r|r|} \hline
MatrixName	& $\alpha_{\scas{GPU,K1}}$	&$\alpha_{\scas{CPU,K1}}$ & $\alpha_{\scas{GPU,K2}}$	&$\alpha_{\scas{CPU,K2}}$\\ \hline
Circuit	&3	&15 &3	&12\\
Economics	&3	&12 &3	&13\\
Epidemiology	&2	&20 &2	&21\\
FEMAccelerator	&3	&12 &3	&12\\
FEMCantilever	&12	&50 &9	&39\\
FEMHarbor	&7	&27 &7	&28\\
FEMShip	&8	&27 &8	&24\\
FEMSpheres	&10	&34 &10	&34\\
\textbf{Heart 3K}	&1	&4 &1	&4\\
\textbf{Heart 5K}	&1	&5 &1	&5\\
\textbf{Heart 30K}	&3	&10 &3	&10\\
Protein	&21	&96 &21	&91\\
QCD	&4	&17 &4	&17\\
Webbase	& $\infty$	& $\infty$ &2	&12\\
WindTunnel	&8	&26 &8	&24\\ \hline
\end{tabular}

%% file: mgpubestvalues.tex

\begin{tabular}{|c|r|r|r|r|r|r|}\hline
MatrixName	&mgpu [1] best	&mgpu [50] best	&mgpu [1200] best\\ \hline
Circuit	&6	&6	&4\\
Economics	&6	&6	&6\\
Epidemiology	&6	&4	&4\\
FEMAccelerator	&6	&6	&4\\
FEMCantilever	&10	&10	&10\\
FEMHarbor	&10	&12	&16\\
FEMShip	&10	&10	&10\\
FEMSpheres	&10	&12	&12\\
Heart 3K	&12	&10	&10\\
Heart 5K	&10	&8	&8\\
Heart 30K	&10	&10	&6\\
Protein	&10	&16	&16\\
QCD	&10	&10	&10\\
Webbase	&6	&4	&4\\
WindTunnel	&10	&10	&10\\ \hline
\end{tabular}

%% file: wpk1bestvalues.tex
\begin{tabular}{|c|r|r|r|r|r|r|}\hline
MatrixName	&K1 [1] best	&K1 [50] best	&K1 [1200] best\\ \hline
Circuit	&64	&64	&64\\
Economics	&64	&64	&64\\
Epidemiology	&128	&128	&128\\
FEMAccelerator	&64	&64	&64\\
FEMCantilever	&64	&64	&64\\
FEMHarbor	&96	&96	&96\\
FEMShip	&96	&96	&96\\
FEMSpheres	&96	&96	&96\\
Heart 3K	&32	&32	&32\\
Heart 5K	&64	&32	&64\\
Heart 30K	&64	&64	&64\\
Protein	&64	&96	&96\\
QCD	&224	&224	&224\\
Webbase	&32	&32	&32\\
WindTunnel	&96	&96	&96\\ \hline
\end{tabular}

%% file: wpk2bestvalues.tex
\begin{tabular}{|c|r|r|r|r|r|r|}\hline
MatrixName & \multicolumn{2}{c|}{K2 [1]} & \multicolumn{2}{c|}{K2 [50]} & \multicolumn{2}{c|}{K2 [1200]} \\
& threshold & blocksize & threshold & blocksize & threshold & blocksize \\ \hline
Circuit	&80	&160	&38	&128	&112	&160\\
Economics	&7	&96	&7	&96	&7	&96\\
Epidemiology	&4	&128	&4	&128	&4	&128\\
FEMAccelerator	&58	&64	&57	&64	&58	&64\\
FEMCantilever	&19	&224	&19	&224	&19	&224\\
FEMHarbor	&128	&96	&132	&96	&139	&96\\
FEMShip	&28	&128	&27	&128	&27	&128\\
FEMSpheres	&47	&128	&81	&96	&81	&96\\
Heart 3K	&7	&256	&7	&256	&6	&192\\
Heart 5K	&10	&192	&13	&96	&6	&192\\
Heart 30K	&21	&64	&17	&64	&17	&64\\
Protein	&172	&96	&167	&96	&171	&96\\
QCD	&39	&96	&39	&224	&39	&128\\
Webbase	&658	&128	&682	&128	&672	&128\\
WindTunnel	&38	&96	&37	&96	&38	&96\\ \hline
\end{tabular}